\documentclass[
journal=jpclcd,
manuscript=letter, 
review]{achemso}

\usepackage{graphicx}
\usepackage{dcolumn}
\usepackage{bm}
\usepackage{xcolor}
\usepackage{amsmath}
\usepackage{mathrsfs}
\usepackage[utf8]{inputenc}
\usepackage[T1]{fontenc}
\usepackage{etoolbox}
\usepackage{subcaption} 
\usepackage[numbers,super]{natbib}
\usepackage[colorlinks=true,citecolor=blue,linkcolor=blue, urlcolor=blue]{hyperref}
\usepackage[font=footnotesize,labelfont=bf, textfont=normal]{caption}
\usepackage[version=3]{mhchem} 
\usepackage{multirow}

\usepackage{pdfpages}

\newcommand{\ket}[1]{\vert #1 \rangle}

\newcommand{\bra}[1]{\langle #1 \vert}

\newcommand{\braket}[1]{\langle #1 \rangle}

\renewcommand{\phi}{\varphi}
\renewcommand{\theta}{\vartheta}
\newcommand{\revHK}[1]{{\color{black}#1}}
\newcommand{\revFR}[1]{{\color{black}#1}}

\newcommand\w[1]{\makebox[2.5em]{$#1$}}

\author{Federico Rossi}
\affiliation{Department of Chemistry, Norwegian University of Science and Technology, NTNU, 7491 Trondheim, Norway}
\author{Eirik F. Kjønstad}
\affiliation{Department of Chemistry, Norwegian University of Science and Technology, NTNU, 7491 Trondheim, Norway}
\author{Sara Angelico}
\affiliation{Department of Chemistry, Norwegian University of Science and Technology, NTNU, 7491 Trondheim, Norway}
\author{Henrik Koch}
\affiliation{Department of Chemistry, Norwegian University of Science and Technology, NTNU, 7491 Trondheim, Norway}

\email{henrik.koch@ntnu.no}

\title[] 
{Generalized coupled cluster theory for ground and excited state intersections}

\begin{document}

\begin{tocentry}
 \includegraphics[width=1.0\textwidth]{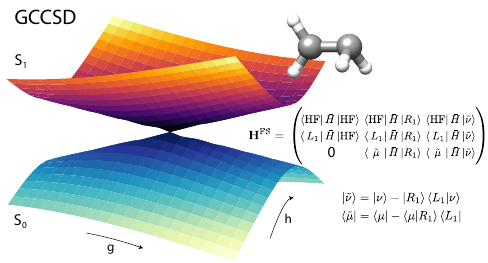}   
\end{tocentry}

\begin{abstract}

Coupled cluster theory in the standard formulation is unable to correctly describe conical intersections among states of the same symmetry. This limitation has restricted the practical application of an otherwise highly accurate electronic structure model, particularly in nonadiabatic dynamics. Recently, the intersection problem among the excited states was fully characterized and resolved. However, intersections with the ground state remain an open challenge, and addressing this problem is our objective here. We present a generalized coupled cluster framework that correctly accounts for the geometric phase effect and avoids bifurcations of the solutions to the ground state equations. Several applications are presented that demonstrate the correct description of ground state conical intersections. We also propose how the framework can be used for other electronic-structure methods.

\end{abstract}

\maketitle

\subsection{Introduction}
Molecular systems with quasi-degeneracies or conical intersections between the ground and excited states present a significant challenge for single-reference coupled cluster methods. Although numerous multireference coupled cluster methods have been proposed over the past forty years, comprehensive assessments indicate that no satisfactory solution has yet been found.\cite{Jeziorski, lyakh2012multireference, Evangelista2018} As a result, coupled cluster methods have not been used to describe ground state conical intersections, even though such degeneracies are critically important to non-radiative relaxation processes found in a wide range of biological and chemical systems.\citep{mai2020molecular}
In this paper, we do not aim to solve the general multireference case, which for instance is needed to describe the dissociation of molecules, limiting
ourselves to the specific case of conical intersections between the ground and excited states.

The description of excited state intersections is also flawed in standard coupled cluster theory,~\citep{hattig2005structure,kohn2007can,kjonstad2017crossing} except for the geometric phase effect.~\citep{williams2023geometric} However, it is now known that the problems associated with excited state intersections (distortion of potential energy surfaces, complex energies, and incorrect topology) can be corrected by enforcing orthogonality relations between the electronic states.
This is the main idea behind similarity constrained coupled cluster (SCC) theory,\citep{kjonstad2017resolving,kjonstad2019orbital} which provides a small correction to standard coupled cluster theory that restores a correct description of conical intersections. This method was recently applied successfully in nonadiabatic dynamics simulations on gas-phase thymine,\citep{kjonstad2024coupled,kjonstad2024unexpected} opening up a range of applications to excited-state relaxation processes.

Recently Kjønstad and Koch\cite{kjonstad2024understanding} demonstrated that the ground state coupled cluster wave function fails to account for the geometric phase encountered when traversing a path around a conical intersection. This leads to divergences in the coupled cluster wave function and results in a multi-valued potential energy surface, where different surfaces can arise depending on the direction of the path taken around the conical intersection. As shown in Ref.~\citenum{kjonstad2024understanding}, these divergences are not only confined to small regions of the potential energy surface but extend throughout the entire configuration space encircling a ground state conical intersection. 

Another complication arises when the Jacobian matrix becomes nearly singular, which occurs near the intersection. This situation defines a bifurcation point\cite{Chow1982} in the amplitude equations, leading to multiple possible solutions. These solutions have been studied in great detail previously.\cite{Adamowicz1994, Kowalski1999, Faulstich2024} However, to the best of our knowledge, a wave function parametrization that eliminates these bifurcations has not been proposed. Each of the multiple solutions may be a reasonable approximation in some regions but completely unphysical in others, and there can be regions of internal coordinate space where the amplitude equations cannot be solved when the amplitudes are restricted to be real. We will show examples of these cases below. 

The study of regions near ground state conical intersections in coupled cluster theory is highly challenging due to the bifurcation of solutions and the breakdowns caused by the phase effect.~\citep{kjonstad2024understanding} This is probably the reason why this area is largely unexplored in the community. The algorithm recently described by Angelico, Kj{\o}nstad, and Koch\cite{angelico2024determining} is therefore an indispensable tool when exploring the configuration space near intersections. This algorithm determines structures on an enveloped seam (also referred to as a tube) that is large enough
to avoid the unphysical regions and sufficiently small to give reliable minimum energy conical intersection structures. These structures are denoted as $\varepsilon$-MECI, where $\varepsilon$ corresponds to the extent of the tube that is wrapped around the seam.

\revHK{Several approaches have been developed to circumvent the ground state intersection problem in electronic structure methods. These methods mainly involve starting from a different state of the system, for instance, a high-spin triplet state followed by a spin-flip (SF) in the equation-of-motion (EOM) treatment (SF-TD-DFT~\citep{Casanova2020SpinflipMI} and EOM-SF-CCSD~\citep{KRYLOV2001}) or starting from a double ionized state and adding electrons back in via the EOM framework (hh-TDA~\citep{Martinex2020_1,Martinez2020} and DEA-EOM-CCSD~\citep{Bartlett1997, Krylov2021}). Other methods, notably in density functional theory (DFT), have been developed that either change the kernel to account for double excited states~\citep{Maitra2004} or by adding an additional double excited configuration to the Hermitian Tamm-Dancoff eigenvalue problem (TDDFT-1D~\citep{subotnik2019}). Ensemble DFT (eDFT) methods have also been applied to strongly correlated systems and extended to treat excited states (SI-SA-REKS \citep{filatov2015spin}). More recently, Schmerwitz \textit{et al.}~\citep{Schmerwitz2023} developed an algorithm to determine saddle points in DFT methods that can be identified with electronic excited states, giving access to ground state intersections. This method is related to the norm-extended optimization scheme for multiconfigurational wave functions~\citep{Jørgensen1984} that also determine saddle point states that are nonorthogonal. }

In this work, we derive a generalized coupled cluster theory (GCC) that avoids all the unphysical behaviors mentioned above. \revHK{The framework corrects the ground state wave function parametrization and does not consider a different state of the system as a starting point.} The complications related to matrix defects in the non-Hermitian eigenvalue problem can be handled using the techniques developed in similarity-constrained coupled cluster theory and will not be discussed here. The application of GCC with singles and doubles excitations (GCCSD) will be illustrated for several molecular systems where the coupled cluster singles and doubles (CCSD) model fails to give a correct description. We will also formulate the GCC2 model, which is a generalization of the well-established CC2 model.\cite{Christiansen1995} Finally, we will propose a procedure to eliminate bifurcations in Hartree-Fock and density functional theory, which in their present formulation are unable to describe ground state conical intersections.\cite{levine2006conical,matsika2021electronic}

\section{Generalized coupled cluster theory}

In standard coupled cluster theory, the electronic wave function is given by
\begin{equation}
     |\mathrm{CC}\rangle = \exp(T)|\mathrm{HF}\rangle,
\end{equation}
where the cluster operator $T$ is expressed in terms of excitation operators $\tau_\mu$ and cluster amplitudes $t_\mu$, such that 
\begin{equation}
    T = \sum_\mu t_\mu \tau_\mu .
\end{equation}
The excitation operators, labeled by the index $\mu$, are defined with respect to a closed-shell Hartree-Fock reference $|\mathrm{HF}\rangle$, and thus the operators commute among one another.
The energy and the amplitudes are determined by projecting the electronic Schrödinger equation on $\{ \bra{\mathrm{HF}}, \bra{\mu} \}$ and we obtain the well-known equations
\begin{align}
    E_0(T) &= \bra{\mathrm{HF}}\bar{H}\ket{\mathrm{HF}} \\
     \Omega_\mu(T) &= \bra{\mu}\bar{H}\ket{\mathrm{HF}} =  0,  \label{eq:omegaeq}
\end{align}
where eq. \ref{eq:omegaeq} is the amplitude equations and $E_0$ is the coupled cluster energy.\cite{helgaker2013molecular} The similarity-transformed Hamiltonian is given by $\bar{H} = \exp{(-T)}H\exp{(T)}$, where $H$ is the electronic Born-Oppenheimer Hamiltonian.\\
The amplitude equations can be expanded in a Taylor series in the following way
\begin{align}
     \Omega_\mu(T+\Delta T) = \Omega_\mu(T) + \sum_\nu A_{\mu \nu} \Delta t_\nu + \tfrac{1}{2}\sum_{\nu \delta} B_{\mu \nu \delta} \Delta t_\nu \Delta t_\delta + \ldots = 0,
     \label{eq:omega_taylor}
\end{align}
where the Jacobian and its derivative are given by 
\begin{align}
     A_{\mu \nu} &= \bra{\mu}[\bar{H}, \tau_\nu ]\ket{\mathrm{HF}} \\
     B_{\mu \nu \delta} &= \bra{\mu}[[\bar{H}, \tau_\nu ], \tau_\delta] \ket{\mathrm{HF}}.
\end{align}
If we assume the Jacobian is diagonalizable ($\mathbf{S}^{-1}\mathbf{AS} = \mathbf{D}$) we may write eq.~\ref{eq:omega_taylor} in the eigenbasis
\begin{align}
     \Omega_k (T+\Delta T) = \Omega_k (T) + \omega_k \Delta t_k + \tfrac{1}{2}\sum_{l m} B_{k l m} \Delta t_l \Delta t_m  +\ldots ,
     \label{eq:omega_eigen}
\end{align}
where
$\Omega_k = (\mathbf{S}^{-1}\Omega)_k$, $\Delta t_k = (\mathbf{S}^{-1}\Delta t)_k$, $\tau_k = (\tau^T \mathbf{S})_k$, and $\omega_k$ is an eigenvalue of $\mathbf{A}$. It is clear that if $\omega_k$ is close to zero, then the higher-order terms in the expansion dominate, giving rise to a bifurcation 
into more than one solution. Depending on the properties
of these higher-order terms, we may obtain different situations with two or more real solutions or even none.
Before we discuss our solution to the bifurcation problem, let us consider the equation of motion eigenvalue problem\cite{Koch_1990_cclr,Stanton1993} for $\bar{H}$
\begin{align}
\mathbf{\bar{H}}_\mathrm{EOM}=\begin{pmatrix}
\bra{\mathrm{HF}}\bar{H}\ket{\mathrm{HF}} & \bra{\mathrm{HF}}\bar{H}\ket{\nu} \\
\bra{\mu} \bar{H}\ket{\mathrm{HF}} & \bra{\mu}\bar{H}\ket{\nu}
\end{pmatrix}=\begin{pmatrix}
E_0 & \eta_\nu \\
0 & A_{\mu \nu}+\delta_{\mu \nu}E_0
\label{eq:eomH}
\end{pmatrix},
\end{align}
where we assume the amplitudes equations have been solved  $(\Omega_\mu = 0)$. From eq.~\ref{eq:eomH} we observe the excitation energies are 
equal to the eigenvalues of the Jacobian $\mathbf{A}$. We introduce the following notation
\begin{align}
    &\mathbf{A}\mathbf{r}_n = \omega_n \mathbf{r}_n \\
    &\mathbf{A}^{\mathrm{T}}\mathbf{l}_n = \omega_n \mathbf{l}_n ,
\end{align}
where the left and right eigenvectors are biorthonormal, that is, $\mathbf{l}_m^T\mathbf{r}_n = \delta_{mn}$. Furthermore, the left and right states are written as
\begin{align}
\ket{R_n} = R_n \ket{\mathrm{HF}} =\sum_\mu \tau_\mu r_{\mu n} \ket{\mathrm{HF}} =\sum_\mu \ket{\mu}r_{\mu n}\\
\bra{L_n} =  \bra{\mathrm{HF}}L_n = \bra{\mathrm{HF}} \sum_\mu \tau^\dag_\mu l_{\mu n} =\sum_\mu l_{\mu n} \bra{\mu},
\end{align}
and we define the biorthogonal projection operators as
\begin{align}
P_n = \ket{R_n}\bra{L_n}.
\end{align}
When $\mathbf{A}$ is diagonalizable this set of projectors is complete, that is
$\sum_n P_n = 1$.
We can now expand the amplitudes in this basis
\begin{align}
    \ket{t} = \sum_n \ket{R_{n}}\braket{L_n|t}
\end{align}
where $\ket{t}=\sum_\nu \ket{\nu}t_\nu$. From the numerical investigation of the CCSD model presented in the Application section, we observe that, for certain solutions to the ground state equations in eq. \eqref{eq:omegaeq}, components of the cluster amplitudes, specifically $\braket{L_n|t}$, can become very large and may even exhibit diverging behavior. We therefore propose to remove these components from the cluster amplitudes and solve the amplitude equations for a restricted set of amplitudes. When projecting out the lowest eigenvector, the effective Jacobian that enters the amplitude equations becomes positive definite and we obtain a convex problem without a bifurcation.
This is the basic idea behind the generalized coupled cluster theory we will outline.

We will initially formulate GCC when projecting only one state. 
The extension to several states is straightforward and is shown in the Supporting Information. We denote this state as $\bra{L_1}$ and $\ket{R_1}$, with the associated Jacobian eigenvalue $\omega_1$, and we introduce the modified projection manifold
\begin{align}
    \ket{\Tilde{\nu}} &= \ket{\nu} - \ket{R_1}\braket{{L_1}|\nu}\\
    \bra{\Tilde{\mu}} &= \bra{\mu} - \braket{\mu|R_1}\bra{L_1}.
\end{align}
We have that $\bra{\Tilde{L}_1} = \sum_\mu l_{\mu 1} \bra{\tilde{\mu}}= 0$ and similarly $\ket{\Tilde{R}_1} = 0 $,  as $\braket{L_1|R_1 }=1$. Working with this set is more convenient, as the block structure of the matrices becomes more transparent. Alternatively, we could have used $\{ \ket{\mathrm{HF}}, \ket{\mathrm{\mu}} \}$, for both the left and right basis.  We now require that the cluster amplitudes do not contain the eigenvector components
\begin{align}
    \ket{t} = \ket{t'} - \ket{R_1}\braket{L_1|t'},
\end{align}
and we determine the remaining amplitudes such that 
\begin{align}
    &\Tilde{\Omega}_\mu = \bra{\Tilde{\mu}}\bar{H}\ket{\mathrm{HF}}=0\\ \label{eq:omega}
    &\mathbf{A}\mathbf{r}_1 = \omega_1 \mathbf{r}_1 \\  \label{eq:jacobi_eigen}
    &\mathbf{A}^{\mathrm{T}}\mathbf{l}_1 = \omega_1 \mathbf{l}_1.
\end{align}
These are coupled sets of equations that we solve using standard techniques. Details about the convergence are presented in the Application section.
It is clear that as we have removed one component from the cluster operator, we cannot in general obtain the full configuration interaction solution for the ground state coupled cluster wave function. However, we may include the projected components in the diagonalization of the similarity-transformed Hamiltonian and thereby obtain the ground and excited states, also in the exact limit. Employing the following left basis $\{ \bra{\mathrm{HF}}, \bra{L_1}, \bra{\Tilde{\mu}} \}$  and right basis $\{ \ket{\mathrm{HF}}, \ket{R_1}, \ket{\Tilde{\nu}} \}$, we obtain the full space eigenvalue equations
\begin{align}
    &\mathbf{H}^{\mathrm{FS}}\mathbf{x}_n = \mathcal{E}_n \mathbf{S}^{\mathrm{FS}}\mathbf{x}_n \label{eq:diagonalize_right} \\
     & \mathbf{y}_n^{\mathrm{T}} \mathbf{H}^{\mathrm{FS}}  = \mathcal{E}_n \mathbf{y}_n^{\mathrm{T}} \mathbf{S}^{\mathrm{FS}},
     \label{eq:FSeigen}
\end{align}
where the matrices are defined as

\begin{equation}
\mathbf{H}^\mathrm{FS}=\begin{pmatrix}
\bra{\mathrm{HF}}\bar{H}\ket{\mathrm{HF}} & \bra{\mathrm{HF}}\bar{H}\ket{R_1} &  \bra{\mathrm{HF}}\bar{H}\ket{\Tilde{\nu}}\\
\bra{L_1}\bar{H}\ket{\mathrm{HF}} & \bra{\mathrm{L}_1}\bar{H}\ket{R_1} & \bra{L_1}\bar{H}\ket{\Tilde{\nu}}\\
\bra{\Tilde{\mu}}\bar{H}\ket{\mathrm{HF}} & \bra{\Tilde{\mu}}\bar{H}\ket{R_1} & \bra{\Tilde{\mu}}\bar{H}\ket{\Tilde{\nu}}
\end{pmatrix}=\begin{pmatrix}
E_0 & \bm{\eta}^T \mathbf{r}_1  & X_\nu\\
\mathbf{l}_1^T \mathbf{\Omega} & E_0 +W_{11} & Y_\nu \\
0 & V_\mu & Z_{\mu \nu}+\delta_{\mu \nu}E_0
\end{pmatrix}
\label{eq:H_fullspace}
\end{equation}

\begin{equation}
\mathbf{S}^\mathrm{FS}=\begin{pmatrix}
1 & \braket{\mathrm{HF}|R_1} & \braket{\mathrm{HF}|\Tilde{\nu}} \\
\braket{L_1|\mathrm{HF}} & 1 & \braket{L_1|\Tilde{\nu}}\\
\braket{\Tilde{\mu}|\mathrm{HF}}  & \braket{\Tilde{\mu}|R_1} & \braket{\Tilde{\mu}|\Tilde{\nu}}
\end{pmatrix}=\begin{pmatrix}
1 & 0 & 0\\
\w0 & \w1 & 0\\
0 & 0 & \braket{\Tilde{\mu}|\Tilde{\nu}}
\end{pmatrix},
\end{equation}

\noindent
and the terms in eq. \ref{eq:H_fullspace} are implicitly defined. Further information can be found in the Supporting Information. In the metric matrix $\mathbf{S}^\mathrm{FS}$, the overlap between the left and right projected bases is given by $\braket{\Tilde{\mu}|\Tilde{\nu}} = \delta_{\mu \nu} - \braket{\mu|R_1}\braket{L_1|\nu}$. Although we can only obtain the full configuration interaction limit in the full space case, it is instructive to consider the $2\times2$ reduced space matrix
\begin{equation}
\mathbf{H}^\mathrm{RS}=\begin{pmatrix}
\bra{\mathrm{HF}}\bar{H}\ket{\mathrm{HF}} & \bra{\mathrm{HF}}\bar{H}\ket{R_1} \\
\bra{L_1}\bar{H}\ket{\mathrm{HF}} & \bra{L_1}\bar{H}\ket{R_1}
\end{pmatrix}=\begin{pmatrix}
E_0 & \bm{\eta}^{\mathrm{T}} \mathbf{r}_1 \\
\mathbf{l}_1^{\mathrm{T}} \mathbf{\Omega} & E_0+W_{11}
\end{pmatrix}.
\end{equation}
In the Application section, we numerically demonstrate that the reduced matrix $\mathbf{H}^\mathrm{RS}$ is an excellent approximation to the corresponding full space eigenvalues. This suggests we can bypass solving the full space eigenvalue equation, reducing the computational cost of the framework.

\revFR{We note that solving the eigenvalue problem for both the reduced matrix and the full matrix can sometimes result in a complex pair of eigenvalues and eigenvectors, which is what is observed in a small region close to the conical intersection (see Fig. \ref{fig:ETH_2D_combined} and Fig. S4). This situation is discussed in Ref.~\citenum{kjonstad2017crossing} for the case of conical intersections among excited states, but the same conclusions can be applied to the problem in Eq. \ref{eq:diagonalize_right}. This final step presents the same characteristics as EOM-CC, where the partial ground state solution obtained in the reduced parameter space acts as the ground state, and the Jacobian matrix is replaced by the full space matrix.}

Before investigating the scaling properties of GCC, a few observations about the geometric phase are in order. In the presence of a conical intersection between the ground and first excited states, the wave functions should exhibit a geometric phase when traversing around the intersection. However, when we exclude the excited state from the cluster amplitudes, we expect that the amplitudes will no longer display a geometric phase. Similarly, the eigenvectors of the Jacobian should also not exhibit any phase. 
On the other hand, the states obtained by diagonalizing the non-Hermitian eigenvalue problem in eq.~\ref{eq:diagonalize_right} will exhibit a correct geometric phase effect, as explained by Williams et al.\citep{williams2023geometric} 
This is confirmed numerically below.

\section{Scaling with system size}

Size extensivity of total energies and size intensivity of excitation energies are two essential properties of coupled cluster theory that ensure scalability to large systems without losing accuracy.\cite{Koch_1990_excitation,helgaker2013molecular} Therefore, we will show that these properties are maintained in GCC. We consider a system composed of two non-interacting subsystems A and B. The total Hamiltonian is the sum of two separate terms $H = H_A + H_B$. When the cluster amplitudes are extensive, $T = T_A + T_B$, we have that $\bar{H} = \bar{H}_A + \bar{H}_B$, where $\bar{H}_X= \exp(-T_X)H_X\exp(T_X)$.  The Hartree-Fock reference is given by the direct product state $\ket{\mathrm{HF}} = \ket{\mathrm{HF_A}}\otimes\ket{\mathrm{HF_B}}$ and the excitation manifold is ordered as $\{ \bra{\mu_A}, \bra{\mu_B}, \bra{\mu_{AB}}\}$.

\noindent
It is well known that when the cluster amplitudes are extensive then the right eigenvectors of the Jacobian in eq.~\ref{eq:jacobi_eigen} are intensive.~\cite{Koch_1990_excitation} Thus we consider a right eigenvector located in system A and denote the excitation operator $R^A_1$. The corresponding left operator $L^A_1$ does not have components in B but can have non-zero elements in the AB part of the operator. We first note that the amplitude equations in eq. \ref{eq:omega} take the form
\begin{align}
    &\Tilde{\Omega}_{\mu_A} = 0 \label{eq:omegatilde_A}\\  
    &\Tilde{\Omega}_{\mu_B} = \Omega_{\mu_B} = 0 \label{eq:omegatilde_B}\\ 
    &\Tilde{\Omega}_{\mu_{AB}} = \Omega_{\mu_{AB}} = 0, \label{eq:omegatilde_AB}
\end{align}
where we have used that $\bra{\Tilde{\mu}_B} = \bra{\mu_B}$ and $\bra{\Tilde{\mu}_{AB}} = \bra{\mu_{AB}}$. We conclude there is a size-extensive solution as the AB part is always zero whenever the cluster amplitudes are extensive. We can now analyze the structure of the full space Hamiltonian matrix

\begin{equation}
\mathbf{H}^\mathrm{FS}=\begin{pmatrix}
\bra{\mathrm{HF}}\bar{H}\ket{\mathrm{HF}} & \bra{\mathrm{HF}}\bar{H}\ket{R_1^A} &  \bra{\mathrm{HF}}\bar{H}\ket{\Tilde{\nu}_{A}} & \bra{\mathrm{HF}}\bar{H}\ket{\nu_{B}} & \bra{\mathrm{HF}}\bar{H}\ket{\Tilde{\nu}_{AB}}\\
\bra{L_1^A}\bar{H}\ket{\mathrm{HF}} & \bra{L_1^A}\bar{H}\ket{R_1^A} & \bra{L_1^A}\bar{H}\ket{\Tilde{\nu}_{A}} &\bra{L_1^A}\bar{H}\ket{\nu_{B}} & \bra{L_1^A}\bar{H}\ket{\Tilde{\nu}_{AB}} \\
\bra{\Tilde{\mu}_{A}}\bar{H}\ket{\mathrm{HF}} & \bra{\Tilde{\mu}_{A}}\bar{H}\ket{R_1^A} & \bra{\Tilde{\mu}_{A}}\bar{H}\ket{\Tilde{\nu}_{A}} &\bra{\Tilde{\mu}_{A}}\bar{H}\ket{\nu_{B}} & \bra{\Tilde{\mu}_{A}}\bar{H}\ket{\Tilde{\nu}_{AB}} \\
\bra{\mu_{B}}\bar{H}\ket{\mathrm{HF}} & \bra{\mu_{B}}\bar{H}\ket{R_1^A} & \bra{\mu_{B}}\bar{H}\ket{\Tilde{\nu}_{A}} &\bra{\mu_{B}}\bar{H}\ket{\nu_{B}} & \bra{\mu_{B}}\bar{H}\ket{\Tilde{\nu}_{AB}} \\
\bra{\mu_{AB}}\bar{H}\ket{\mathrm{HF}} & \bra{\mu_{AB}}\bar{H}\ket{R_1^A} & \bra{\mu_{AB}}\bar{H}\ket{\Tilde{\nu}_{A}} &\bra{\mu_{AB}}\bar{H}\ket{\nu_{B}} & \bra{\mu_{AB}}\bar{H}\ket{\Tilde{\nu}_{AB}} \\
\end{pmatrix}.
\end{equation}
where we have used that $\ket{\Tilde{\nu}_B} = \ket{\nu_B}$. Using eqs. \ref{eq:omegatilde_A}-\ref{eq:omegatilde_AB} we may show that
\begin{align}
    &\bra{\mu_{B}}\bar{H}\ket{\mathrm{HF}} = \bra{\mu_{B}}\bar{H}\ket{R_1^A} = \bra{\mu_{B}}\bar{H}\ket{\Tilde{\nu}_{A}} = 0 \\
    &\bra{\mu_{AB}}\bar{H}\ket{\mathrm{HF}} = \bra{\mu_{AB}}\bar{H}\ket{R_1^A} = \bra{\mu_{AB}}\bar{H}\ket{\Tilde{\nu}_{A}} =0,
\end{align}
and this leads to the following block structure of the matrix
\begin{equation}
\mathbf{H}^\mathrm{FS}=\begin{pmatrix}
\mathbf{H}^{\mathrm{FS}}_{A,A} & \mathbf{H}_{A,B} & \mathbf{H}_{A,AB}\\
\mathbf{0} & \mathbf{H}_{B,B} & \mathbf{H}_{B,AB}\\
\mathbf{0} & \mathbf{H}_{AB,B} & \mathbf{H}_{AB,AB}  \label{eq:fullspace_A}
\end{pmatrix}.
\end{equation}
The full space matrix in the A system is given by
\begin{equation}
\mathbf{H}^\mathrm{FS}_{A,A} =\begin{pmatrix}
\bra{\mathrm{HF}}\bar{H}\ket{\mathrm{HF}} & \bra{\mathrm{HF}}\bar{H}\ket{R_1^A} &  \bra{\mathrm{HF}}\bar{H}\ket{\Tilde{\nu}_{A}} \\
\bra{L_1^A}\bar{H}\ket{\mathrm{HF}} & \bra{L_1^A}\bar{H}\ket{R_1^A} & \bra{L_1^A}\bar{H}\ket{\Tilde{\nu}_{A}}\\
\bra{\tilde{\mu}_{A}}\bar{H}\ket{\mathrm{HF}} & \bra{\tilde{\mu}_{A}}\bar{H}\ket{R_1^A} & \bra{\tilde{\mu}_{A}}\bar{H}\ket{\Tilde{\nu}_{A}} \label{eq:hfs_AA}
\end{pmatrix},
\end{equation}
and the other matrix elements in eq. \ref{eq:fullspace_A} are implicitly defined. Further analysis of this matrix reveals that
\begin{equation}
    \mathbf{H}^\mathrm{FS}_{A,A} = \mathbf{H}^A_{A,A} +  \mathbf{I}_{A,A}E_0^B,
\end{equation}
where $\mathbf{H}^A_{A,A}$ is obtained from eq. \ref{eq:hfs_AA} replacing $\bar{H}$ with $\bar{H}_A$. Thus we have shown that the total energies in A are size-extensive and the excitation energies are size-intensive as the diagonal elements are shifted with ground state energy of the B system.

\noindent
We now consider size-extensivity in the B system and introduce another system C that is non-interacting with both A and B. We consider the matrix
\begin{equation}
\begin{pmatrix}
\mathbf{H}_{B,B} & \mathbf{H}_{B,AB} & \mathbf{H}_{B,C} & \mathbf{H}_{B,AC}\\
\mathbf{H}_{AB,B} & \mathbf{H}_{AB,AB} & \mathbf{H}_{AB,C} & \mathbf{H}_{AB,AC}\\
\mathbf{H}_{C,B} & \mathbf{H}_{C,AB} & \mathbf{H}_{C,C} & \mathbf{H}_{C,AC}\\
\mathbf{H}_{AC,B} & \mathbf{H}_{AC,AB} & \mathbf{H}_{AC,C} & \mathbf{H}_{AC,AC}
\end{pmatrix} =
\begin{pmatrix}
\mathbf{H}_{B,B} & \mathbf{H}_{B,AB} & \mathbf{0} & \mathbf{0}\\
\mathbf{H}_{AB,B} & \mathbf{H}_{AB,AB} & \mathbf{0} & \mathbf{0}\\
\mathbf{0} & \mathbf{0} & \mathbf{H}_{C,C} & \mathbf{H}_{C,AC}\\
\mathbf{0} & \mathbf{0} & \mathbf{H}_{AC,C} & \mathbf{H}_{AC,AC}
\end{pmatrix},
\label{eq:block_matrix_BC}
\end{equation}
and we observe that all the new coupling blocks are zero, giving size-extensivity and size-intensity in the B and C systems. We should point out, that the $E_0^A$ is not equal to the ground state energy obtained from eq. \ref{eq:hfs_AA}, since  $\Omega_{\mu_A} \neq 0$. This also implies that the coupling blocks, for instance $\mathbf{H}_{AB,B}$, modify the excitation energies in B and C. In the case where we have sufficiently high excitations to obtain full configuration interaction (FCI) in both A and B or in A and C, the modifications from the coupling blocks provide the FCI excitation energies in B or C. These errors are tiny and do not scale with the overall system size because of the block structures outlined above. Numerical examples will be presented in the Application section. We note that if Jacobian eigenvectors from different subsystems are projected simultaneously, the resulting energies will not be extensive and the errors will scale with the number of subsystems projected.

\begin{figure}
    \centering
    \includegraphics[width=0.97\linewidth]{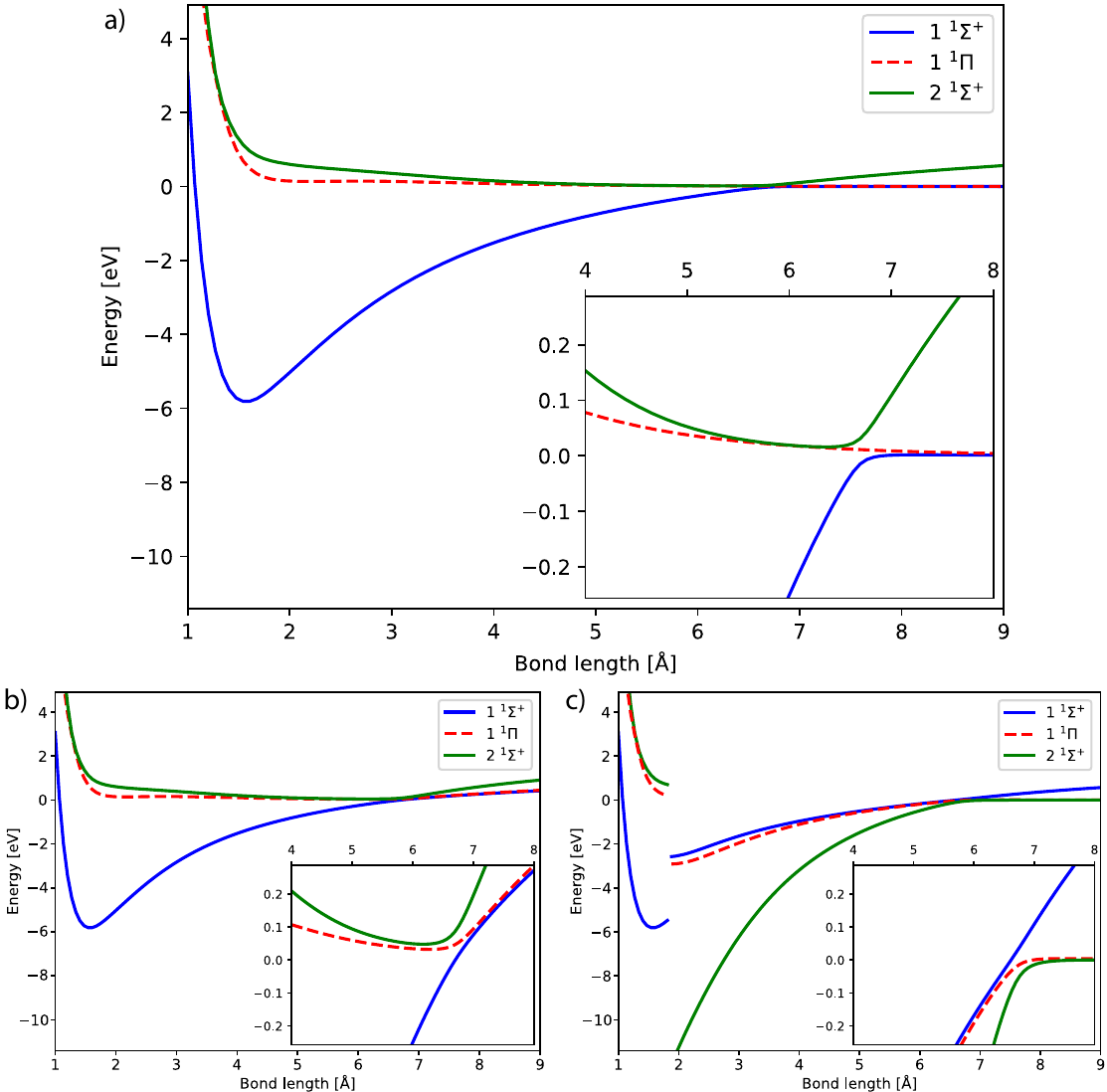}
    \caption{\footnotesize a) GCCSD energy levels of LiF at different interatomic distances while projecting 3 states. The CCSD results are reported in b) starting from 1.0 \AA \,and increasing the interatomic distance and in c) starting from 9.0 \AA\, and decreasing the interatomic distance. The inset shows a close-up view of the avoided crossing.}
    \label{fig:LiF_combined}
\end{figure}

\section{Applications}
The GCCSD framework, for an arbitrary number of projected states, was implemented in a local development version of the $e^T$ program \cite{folkestad20201} and used in all the calculations reported here. Initially, we will consider the lithium fluoride molecule, a standard benchmark system for multireference coupled cluster methods.\cite{Köhn_2016} At a bond distance of about $6.75$ Å the molecule has an avoided crossing \cite{wan2015low} between the $2 ^1\Sigma^{+}$ state and the ground state $1^1 \Sigma^{+}$. We employ an aug-cc-pVDZ basis and find two solutions to the CCSD ground state equations.
The CCSD solution (Fig. \ref{fig:LiF_combined}b), obtained when starting at distances close to the equilibrium bond length ($1.5$ \AA), shows an unexpected increasing diverging trend in the energies of all 3 states when extended past the avoided crossing  and restarting the algorithm from the previous point. Another CCSD solution is found (Fig. \ref{fig:LiF_combined}c) when starting at a long bond distance ($9.0$ \AA). We observe a decreasing diverging trend in the energies when extended to shorter distances. The solution eventually changes discontinuously to the other CCSD solution at about $2.0$ \AA. On the other hand, the GCCSD curves shown in Fig. \ref{fig:LiF_combined}a are continuous for the whole range of bond distances.  Thus, GCCSD bridges the CCSD solution for distances shorter than the avoided crossing with the other CCSD solution beyond the avoided crossing point. In doing so, GCCSD corrects the divergent behavior seen in the two individual CCSD solutions.

\begin{figure}[!htb]
    \centering
    \includegraphics[width=1.0\textwidth]{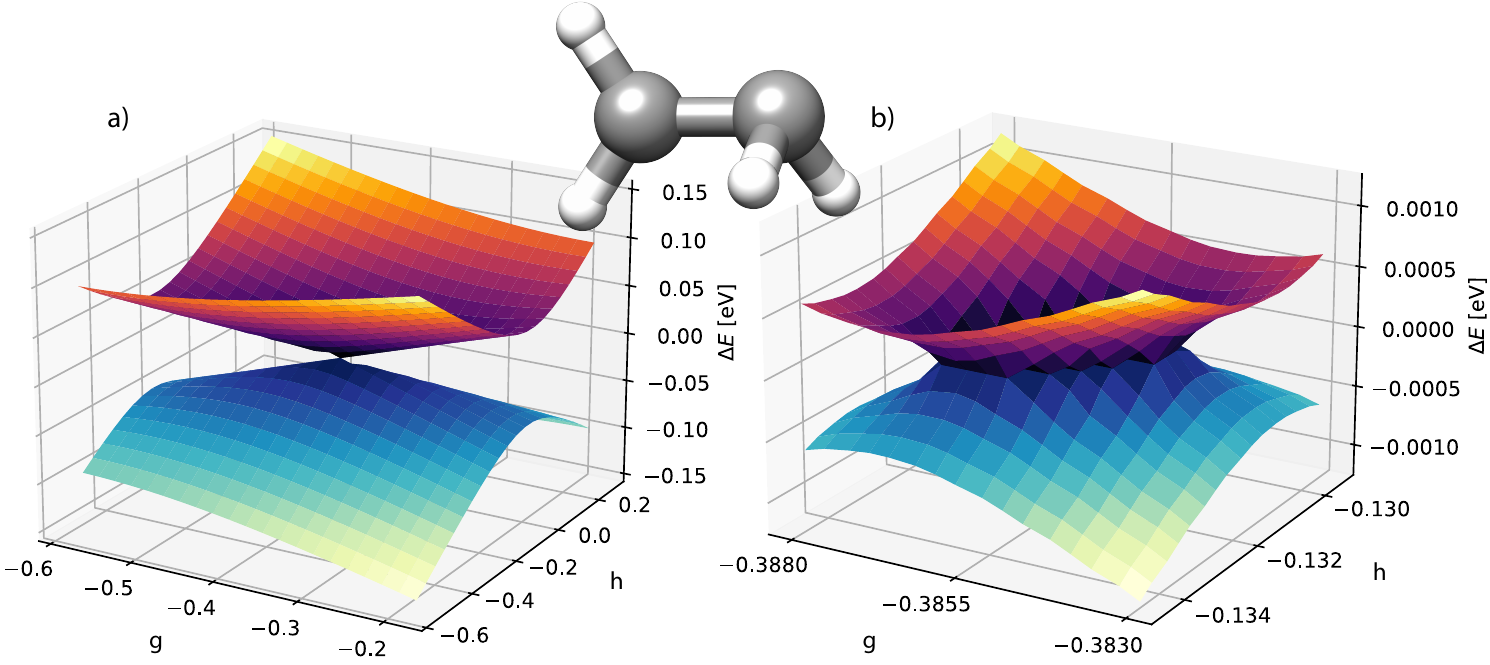}
    \caption{\footnotesize \revFR{(a)} The GCCSD potential energy surfaces of $\mathrm{S}_0$ and $\mathrm{S}_1$ in ethylene. \revFR{(b) A detailed view of the region close to the conical intersection.} The basis is aug-cc-pVDZ and \revFR{for each point} the energies are plotted relative to the average energy $\frac{1}{2}(E_0+E_1)$. \revFR{A plot of the same region, showing the absolute energies of the two states in Hartree can be found in the Supporting Information. }The $\varepsilon$-MECI structure is shown in the middle, which corresponds to the geometry at the point $(\mathrm{g,h}) = (0.0,0.0)$. }
    \label{fig:ETH_2D_combined}
\end{figure}

\begin{figure}[!htb]
    \centering
    \includegraphics[width=0.5\textwidth]{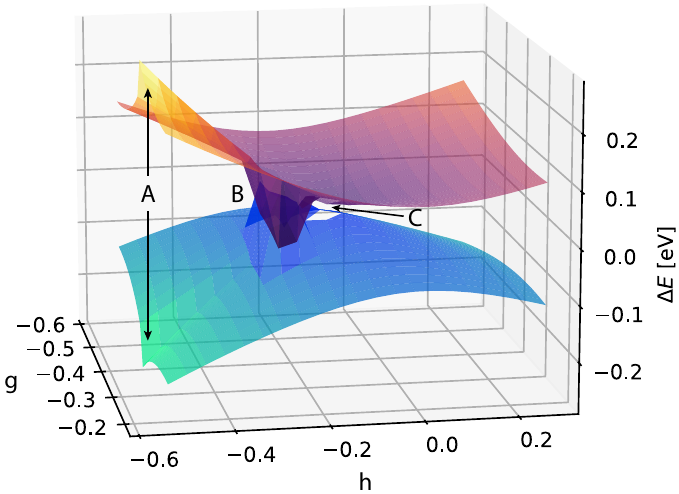} 
    \caption{\footnotesize The CCSD potential energy surfaces of $\mathrm{S}_0$ and $\mathrm{S}_1$ in ethylene. The basis is aug-cc-pVDZ and \revFR{for each point} the energies are plotted relative to the average energy $\frac{1}{2}(E_0+E_1)$. There are three notable regions: in \textbf{A}, a mismatch in energies appears due to the phase effect; 
    in \textbf{B}, a new set of flipped solutions is obtained, characterized by negative excitation energies; and in \textbf{C}, the region where we were not able to converge the ground state equations.
    }
    \label{fig:ETH_2d_plot_ccsd}
\end{figure}

We now consider a ground state conical intersection in ethylene. We will start from the CCSD $\varepsilon$-MECI using an aug-cc-pVDZ basis to explore the potential energy surfaces close to a conical intersection. The $\varepsilon$-MECI structure is reported
by Angelico et al. \cite{angelico2024determining} and is shown in Fig.~\ref{fig:ETH_2D_combined}. We further employ the $\mathbf{g}$ and $\mathbf{h}$ vectors calculated at the $\varepsilon$-MECI geometry using the CCSD algorithm described in Ref.\citenum{Kjønstad2023_1}. Further computational details are reported in the Supporting Information.

In Fig.~\ref{fig:ETH_2D_combined}a we show a GCCSD conical intersection between $\mathrm{S}_0$ and $\mathrm{S}_1$. When we zoom in on the intersection region, shown in Fig.~\ref{fig:ETH_2D_combined}b, we observe a small defective area where the full space matrix has complex energies. Aside from a minor defect, the intersection exhibits the correct conical shape. This is the expected behavior and is explained in earlier works.~\cite{kjonstad2017crossing} However, when compared to a typical CCSD calculation in the branching plane, the difference is striking, as shown in Fig.~\ref{fig:ETH_2d_plot_ccsd}, which clearly illustrates the breakdown of the CCSD model. Here we have mismatches in energies due to problems describing the geometric phase,\citep{kjonstad2024understanding} the bifurcation of the solution resulting in regions with negative excitation energies, and a region where we were not able to converge the ground state equations.

\begin{figure}[!htb]
    \centering
    \includegraphics[width=0.97\textwidth]{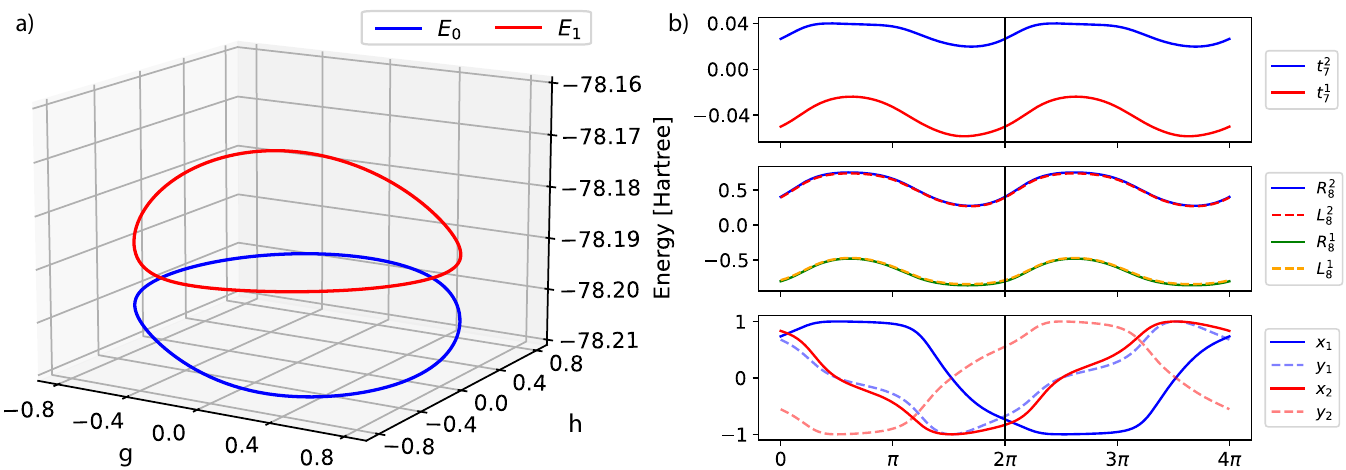}
    \caption{\footnotesize The GCCSD potential energy curves of $\mathrm{S}_0$ and $\mathrm{S}_1$ in ethylene (a), when traversing on a circle around the conical intersection, using aug-cc-pVDZ. In (b), some selected GCCSD parameter values are reported for a $4\pi$ rotation around the intersection. Starting from the top, the two largest components of the cluster amplitudes, the two largest components of the left and right eigenvectors, and finally the 2 eigenvectors of the reduced matrix, where $(x_1, y_1)$ refer to $\mathrm{S}_0$ and $(x_2, y_2)$ to $\mathrm{S}_1$.}
    \label{fig:ETH_circle_combined}
\end{figure}

\begin{figure}[!htb]
    \centering
    \includegraphics[width=0.97\textwidth]{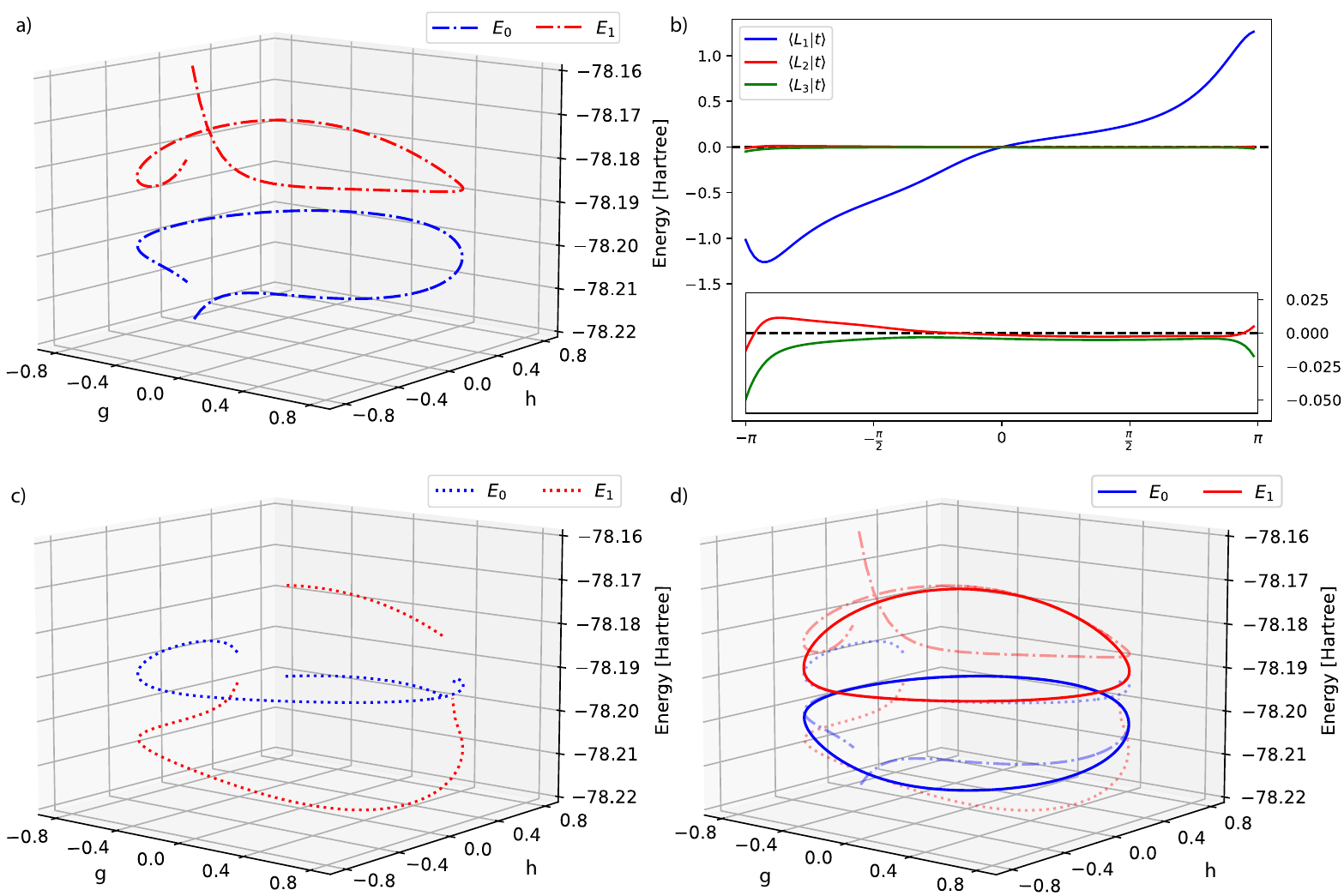}
    \caption{\footnotesize The CCSD potential energy curves of $\mathrm{S}_0$ and $\mathrm{S}_1$ in ethylene, when traversing on a circle around the conical intersection, using aug-cc-pVDZ. The two sets of curves in (a) and (c) have been obtained starting from opposite points, $(g,h) = (0, \pm 0.8)$ respectively, and restarting from the previous point when mapping out a half circle in both directions. In (b) is shown the projection of the amplitudes on the 3 lowest left eigenvectors for the solution in (a) when rotating $\pm\pi$. In (d) the same curves are plotted together with the GCCSD energies (solid line) shown in Fig.~\ref{fig:ETH_circle_combined}a.}
    \label{fig:ETH_circle_multi_combined}
\end{figure}

We now investigate the geometric phase effect in GCCSD. In Fig.~\ref{fig:ETH_circle_combined}a we have mapped out the potential energy curves of $\mathrm{S}_0$ and $\mathrm{S}_1$ when traversing a circle around the conical intersection. The Supporting Information provides more details on how these plots are generated. We observe that both curves are continuous without any artifacts. 
In Fig.~\ref{fig:ETH_circle_combined}b
we examine the behavior of the parameters during a $4\pi$ rotation. We see that the cluster amplitudes and the eigenvectors of the Jacobian are unchanged after a
$2\pi$ rotation. The eigenvectors of the reduced space matrix display the geometric phase effect and change sign at $2\pi$ and return to the original value after $4\pi$. Thus we conclude that GCC correctly accounts for the geometric phase effect. 

In Fig.~\ref{fig:ETH_circle_multi_combined} we present the
corresponding CCSD curves. In Figs.~\ref{fig:ETH_circle_multi_combined}a and ~\ref{fig:ETH_circle_multi_combined}c we have mapped out the two different solutions that are obtained due to the bifurcation in the ground state equations. The solution shown in Fig.~\ref{fig:ETH_circle_multi_combined}a encounters problems as we approach the point where the weight of the Hartree-Fock reference becomes zero -- this situation is discussed in detail in Ref.~\citenum{kjonstad2024understanding}. The other solution is shown in Fig.~\ref{fig:ETH_circle_multi_combined}c where the ground state solution is close to $\mathrm{S}_1$ and the excitation energy is negative. As we traverse the circle, the solution encounters a region where 
the equations do not converge, followed by a change to the solution shown in Fig.~\ref{fig:ETH_circle_multi_combined}a. In Fig.~\ref{fig:ETH_circle_multi_combined}b we show the different components of the CCSD amplitudes and observe the diverging behavior of the component along $L_1$. In Fig.~\ref{fig:ETH_circle_multi_combined}d we have shown GCCSD together with the two CCSD solutions. We observe that CCSD is reasonably close to GCCSD in some regions of the circle but is far away in others.

\begin{table}[h]
\caption{\footnotesize Two lowest excitation energies of ethylene in 3 similar but different geometries (A, B and C). For all GCCSD calculations, only one state is projected which is always localized on A. In the two last columns, the system is composed of multiple ethylene molecules in different geometries and shifted by 1000 Å so that they do not interact with each other. All calculations are performed with aug-cc-pVDZ, with convergence threshold of $1\cdot 10^{-10}$.}
\begin{tabular}{lcccc}
\hline
\hline
System & CCSD X & GCCSD A &  GCCSD AB & GCCSD ABC\\
\hline
$\Delta E^A_1$  &  0.0168 318 946  & 0.0168 301 978 & 0.0168 301 976   &   0.0168 301 975  \\
$\Delta E^B_1$  &  0.0172 860 929  & -              & 0.0172 863 677   &   0.0172 863 676  \\
$\Delta E^C_1$  &  0.0177 393 431  & -              & -                &   0.0177 396 178  \\
$\Delta E^A_2$  &  0.1511 668 829  & 0.1511 672 134 & 0.1511 672 136   &   0.1511 672 135  \\
$\Delta E^B_2$  &  0.1512 823 538  & -              & 0.1512 826 298   &   0.1512 826 296  \\
$\Delta E^C_2$  &  0.1514 009 837  & -              & -                &   0.1514 012 593  \\
\hline
\hline
\end{tabular}
\label{tab:ethylene_ABC}
\end{table}

To demonstrate the size-intensivity of GCCSD excitation energies, we consider a system of three ethylene molecules in similar but distinct geometries: A, B, and C. The molecules are spaced 1000 Å apart, ensuring they do not interact with each other. In the first column of Table \ref{tab:ethylene_ABC}, the first two CCSD excitation energies are reported for the 3 isolated molecules. The second column reports the GCCSD excitation energies on system A alone where we project the first excited state. When system B is included (see the third column), the projected state is localized on A, and system A's excitation energies are unaffected. Excitations in B are only slightly modified, due to the coupling block $\mathbf{H}_{AB,B}$ in eq. \ref{eq:fullspace_A}. When an additional non-interacting ethylene molecule C is added, excitation energies in both A and B are unaffected, as expected from the block structure in eq. \ref{eq:block_matrix_BC}. In the Supporting Information, we report additional cases that illustrate the scaling properties.

\begin{table}[hbt!]
\caption{\footnotesize Energies and information about convergence for one ethylene molecule. Both CCSD and GCCSD calculations determine three excited states with a threshold of $1.0\cdot 10^{-10}$. For GCCSD, the first excited state is projected. 
The numbers of iterations reported are, from top to bottom: the number of DIIS iterations to solve the CCSD ground state equations, and the total number of Davidson's solver iterations to obtain the right and left eigenvectors of the CCSD Jacobian. 
For GCCSD the number of DIIS iterations to solve the cluster amplitude equations. The number of iterations to find the right and left eigenvectors of the GCCSD Jacobian. Finally, the number of Davidson's iterations to determine the right eigenvectors of the full matrix. The timings are wall times in seconds for the entire calculation on an Intel Xeon Gold 6342 using 24 cores. }
\label{tab:ETH_convergence}
\centering
\begin{tabular}{ccrcc}
\hline
\hline
\multirow{3}{*}{System} & \multicolumn{2}{c}{\multirow{3}{*}{Energy}} &  \multirow{3}{*}{Iterations}    & \multirow{3}{*}{Time} \\
                        &                         &             & \\
                        &                         &            & \\
\hline
\multirow{3}{*}{CCSD} & $E_0$      &    -78.1978 872 879 &   33 & \multirow{3}{*}{57.0s}\\
                      & $\omega_1$ &    0.0168 318 946   &  33+33 &\\
                      & $\omega_2$ &    0.1511 668 829   &   &  \\
\hline
\multirow{5}{*}{GCCSD} & $E_0^\mathrm{FS}$       & -78.1978 872 787    &   25 & \multirow{5}{*}{96.3s}\\
                       & $E_0^\mathrm{RS}$       & -78.1978 872 810 &    &\\
                       & $\omega_1^\mathrm{FS}$  & 0.0168 301 977  &    101+104 &\\
                       & $\omega_1^\mathrm{RS}$  & 0.0168 302 023 &       & \\
                       & $\omega_2^\mathrm{FS}$  & 0.1511 672 134 &       27 &\\
\hline
\hline
\end{tabular}
\end{table}

In concluding our investigation of ethylene, we discuss the convergence properties of the framework. We use the ethylene A geometry from the extensivity study above, given in the Supporting Information. In Table~\ref{tab:ETH_convergence} we report the energies for the ground state and the two first excited states. The reduced space numbers are almost identical to the full space and in turn, these are similar to the CCSD numbers.  Even though the number of iterations indicates a factor of $2.5$ in computational cost between GCCSD and CCSD, the wall times show only a factor of $1.7$. This is due to relatively fewer linear transformations in GCCSD compared to CCSD per iteration.

\begin{figure}[!htb]
\centering
\includegraphics[width=\textwidth]{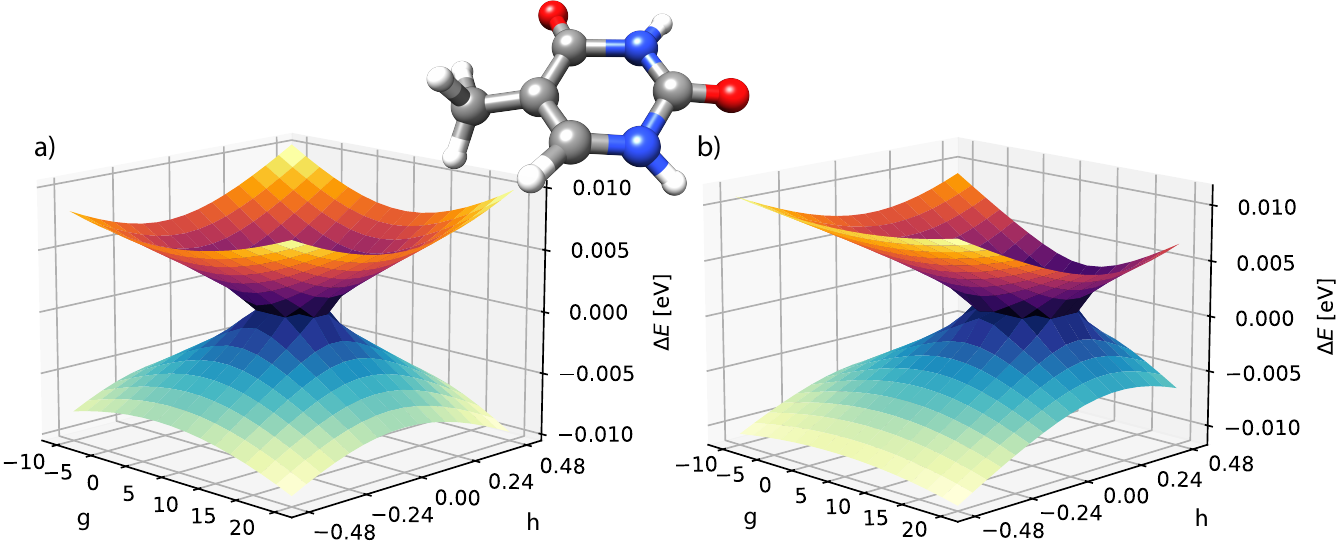}
\caption{\footnotesize CCSD (a) and GCCSD (b) potential energy surfaces of $\mathrm{S}_1$ and $\mathrm{S}_2$ in thymine with cc-pVDZ basis. All energies are plotted in eV, relative to the average $\frac{1}{2}(E_1+E_2)$ \revFR{for each point}. The conical intersection structure is shown in the middle, which corresponds to the geometry at the point $(\mathrm{g,h}) = (0.0,0.0)$.}
\label{fig:THYMINE_combined}
\end{figure}

To demonstrate the GCCSD method when projecting out several states we consider the $\mathrm{S_1}/\mathrm{S_2}$ conical intersection in thymine. We project out the eigenvectors for the two lowest eigenvalues of the Jacobian. This means the reduced space matrix is a $3\times3$ matrix. We use an initial structure together with $\mathbf{g}$ and $\mathbf{h}$ vectors that were determined in Ref.~\citenum{kjonstad2024unexpected}. These are reported in the Supporting Information. In Fig.~\ref{fig:THYMINE_combined} we show the $\mathrm{S}_1$ and $\mathrm{S}_2$ potential energy surfaces for CCSD and GCCSD. \revFR{The surfaces look very similar and they both} have a defect that can be removed using a similarity-constrained transformation\revFR{, as shown in Fig. S3 for CCSD}. The center of the defect is slightly shifted when comparing the two. 

\begin{figure}[!htb]
    \centering
    \includegraphics[width=1.0\textwidth]{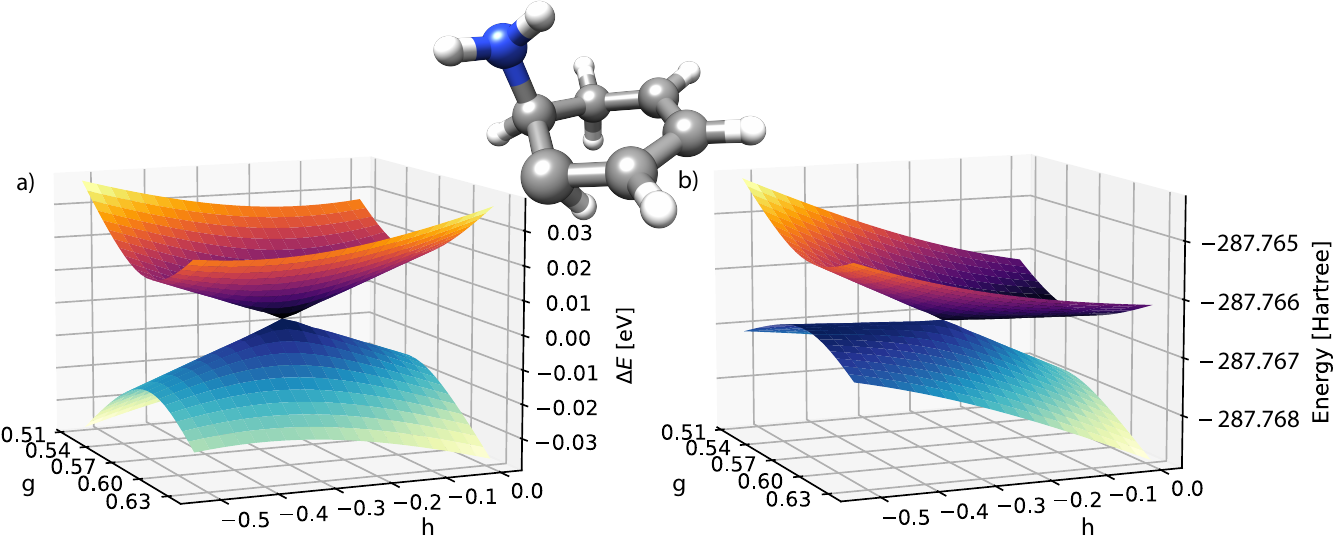}
    \caption{\footnotesize The GCCSD potential energy surfaces of $\mathrm{S}_0$ and $\mathrm{S}_1$ in 2,4-cyclohexadien-1-ylamine with cc-pVDZ. In (a) the energies \revFR{for each point} are plotted in eV relative to the average energy $\frac{1}{2}(E_0+E_1)$ whereas in (b) the total energies are shown in Hartree. The initial structure is shown in the middle, which corresponds to the geometry at the point $(\mathrm{g,h}) = (0.0,0.0)$.}
    \label{fig:1AmCHD}
\end{figure}

As a final example, we show a $\mathrm{S}_0$/$\mathrm{S}_1$ intersection in 2,4-cyclohexadien-1-ylamine. The geometry was obtained from the database in Ref.~\citenum{database_polyene} and the structure is shown in Fig.~\ref{fig:1AmCHD}. The $\mathbf{g}$ and $\mathbf{h}$ vectors are calculated using the CCSD algorithm described in Ref.\citenum{Kjønstad2023_1}. In the same figure, we show the two potential energy surfaces in two different representations. The conical shape of the intersection is visible in both plots. \revFR{Again, when zooming in on the intersection region, a small defective area is observed (see Fig. S2 in the Supporting Information).}

\section{Further perspectives}

The framework presented above can be extended to other electronic structure methods with similar problems accounting for the geometric phase associated with conical intersections with the ground state. For \revHK{example}, time-dependent Hartree-Fock (TDHF) and time-dependent density functional theory (TDDFT) cannot describe the intersection because the ground and excited states are decoupled.\citep{levine2006conical,Curchod2024} 
Other methods such as the algebraic diagrammatic construction hierarchy (ADC)\cite{Schirmer1982, Dreuw2015} also do not include coupling to the ground state. \revHK{The ADC framework relies on Møller-Plesset perturbation theory for the ground state wave function and as shown in Ref.~\citenum{kjonstad2024understanding} the perturbation series converge to an excited state close to the conical intersection. Recently, Taylor et al.~\citep{Curchod2023} discussed the failure of the ADC(2) model at ground state intersections.}

\begin{figure}[!htb]
    \centering
    \includegraphics[width=0.5\textwidth]{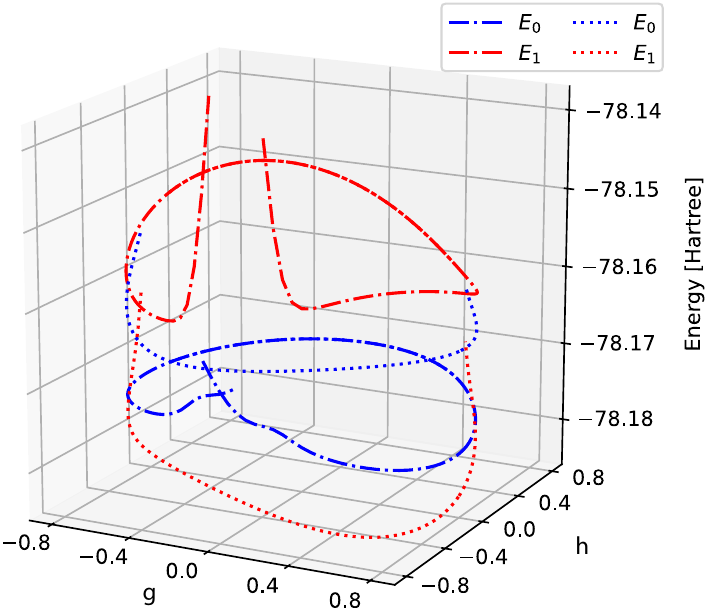} 
    \caption{\footnotesize The CC2 ground state and first excited state energies when traversing a circle around the conical intersection in ethylene, using aug-cc-pVDZ.  The two sets of curves have been obtained starting from opposite points, $(g,h) = 0, \pm 0.8$ for the dash-dot and dotted lines respectively, and restarting half circle in both directions. The red dash-dot line is cut to limit the $z$-dimension, but a crossing between the two ends is present, similar to what can be seen for the blue dash-dot line.}
    \label{fig:ETH_circle_cc2}
\end{figure}

The entire coupled cluster hierarchy is generally affected by the geometric phase issues highlighted \revHK{in Ref.~\citenum{kjonstad2024understanding}}. Here we explicitly mention the CC2 method,\cite{Christiansen1995} which is frequently used for calculating excitation energies of large molecules, due to the favorable balance between computational cost and accuracy compared to CCSD.
The CC2 model is viewed as the best alternative to second-order Møller-Plesset theory as both ground and excited states are available with the same computational cost and would therefore be an excellent candidate for nonadiabatic dynamics.
However, the CC2 method also fails to describe ground state conical intersections, as shown in Figure \ref{fig:ETH_circle_cc2} for the ethylene molecule. Using the GCC framework above, we may extend CC2 to GCC2 by considering the CC2 Hamiltonian matrix

\begin{equation}
\mathbf{H}^\mathrm{CC2}=\begin{pmatrix}
E_0 &\bm{\eta}^\mathrm{T} \\
\mathbf{\Omega}^\mathrm{CC2} & \mathbf{A}^\mathrm{CC2}+\mathbf{I}E_0
\end{pmatrix}.
\end{equation}
\noindent
The explicit expressions for $\mathbf{\Omega}^\mathrm{CC2}$ and $\mathbf{A}^\mathrm{CC2}$ can be found in Ref.\citenum{Christiansen1995}. We will report on the implementation of GCC2 elsewhere together with a detailed benchmarking of the method.

We now outline how the conical intersection problem could be solved for closed-shell Hartree-Fock. Due to the closed shell form of the Hartree-Fock wave function, a rotation of the orbitals cannot change the overall sign of the wave function. This implies that Hartree-Fock theory cannot detect an intersection with the excited state. However, as discussed in detail by Helgaker \textit{et. al},\cite{helgaker2013molecular} the eigenvalues of the electronic Hessian may become small and this can lead to bifurcations in the Hartree-Fock solution. Using the ideas from GCC we may remove these bifurcations.
For this purpose, we parameterize the wave function in terms of orbital rotations
\begin{align}
    \ket{\mathrm{HF}} = \exp(\sum_{ai} \kappa_{ai}E^-_{ai}) \ket{\Phi_0} ,\label{eq:Hartree-Fock}
\end{align}
where $E_{ai}^- = E_{ai} - E_{ia}$, $a,b$ and $i,j$ labels virtual and occupied orbitals, respectively, and  $\ket{\Phi_0}$ is the initial wave function. The gradient and Hessian are given by 
\begin{align}
   E^{(1)}_{ai} &= \bra{\mathrm{HF}}[H,E^-_{ai}]\ket{\mathrm{HF}}\\
   E^{(2)}_{ai,bj} &= \frac{1}{2}(1+P_{ai,bj})\bra{\mathrm{HF}}[[H,E^-_{ai}],E^-_{bj}]\ket{\mathrm{HF}} ,\label{eq:Hessian_HF}
\end{align}
where $P_{ai,bj}$ permutes $ai$ and $bj$ indices. We may now remove the Hessian eigenvector $\mathbf{r}$ from the $\bm{\kappa}$ vector, which is associated with a small eigenvalue that gives rise to bifurcations. This leads to a coupled set of equations that must be solved in the same way as for GCC. The removed component may be included through diagonalization of the matrix
\begin{equation}
\mathbf{H}^\mathrm{RS}=\begin{pmatrix}
\bra{\mathrm{HF}}H\ket{\mathrm{HF}} & \bra{\mathrm{HF}}H\ket{\mathrm{R}} \\
\bra{\mathrm{R}}H\ket{\mathrm{HF}} & \bra{\mathrm{R}}H\ket{\mathrm{R}} 
\end{pmatrix},\label{eq:Hartree-RS}
\end{equation}
where $\ket{\mathrm{R}} = \sum_{ai} \ket{ai} r_{ai}$. The obtained solutions will be able to account for the geometric phase and describe a conical intersection between the two states. We should note that the energies will depend on the initial determinant $\ket{\Phi_0}$, as we remove components from the orbital rotation operator. Furthermore, the Hartree-Fock wave function in eq. \ref{eq:Hartree-Fock} will not be exact for one electron. Therefore it would be more appropriate to include all single excitations in eq. \ref{eq:Hartree-RS} and consider a full space matrix as in GCC.

In passing, we note that as the eigenvectors of the Hessian in eq. \ref{eq:Hessian_HF} are size-intensive then the resulting wave functions will be extensive. They can therefore serve as a reference for a coupled cluster model that extends the GCC framework presented here. Needless to say, the framework can also be used for Tamm-Dancoff TDDFT.\cite{matsika2021electronic,levine2006conical}

\section{Conclusions}

In this paper, we have presented a coupled cluster framework capable of describing intersections with the ground state while accounting for the geometric phase effect and eliminating bifurcations in the ground state equations. This development paves the way for studying nonadiabatic dynamics among excited states
with the possibility of also describing
relaxation to the ground state. Indeed, we have already established techniques that can remove defects in the non-Hermitian eigenvalue problem. This similarity-constrained coupled cluster theory is directly transferable to GCC and analytical molecular gradients and derivative couplings can be calculated with minor modifications to the existing developments for CCSD and SCCSD.

Another aspect of the reported development concerns coupled cluster theory itself. In Ref.~\citenum{kjonstad2024understanding}, we have recognized the need to move away from assuming that the coupled cluster ground state wave function is exact. Even the full coupled cluster wave function is not well-defined in an $(N-1)$ dimensional configuration space due to intermediate normalization. Therefore, instead of imposing exactness, we should focus on  
making sure
that the coupled cluster wave function is well-behaved. 
The exact limit may instead be achieved by subsequent diagonalization of the similarity-transformed Hamiltonian matrix, which at the same time allows for a correct description of conical intersections with the ground state. 
As we have discussed, the approach also extends to other electronic structure theories, such as Hartree-Fock and density functional theory, opening up
an interesting perspective for future developments in electronic structure theory.

\section*{Author contributions}
HK and FR conceived the GCC framework. HK supervised the project. FR and EFK developed the implementation in eT. FR developed the implementation in Julia and performed all the calculations. SA calculated the $\varepsilon$-MECIs, $\mathbf{g}$ and $\mathbf{h}$ vectors. HK and FR wrote the first draft of the paper. All authors analyzed the results and revised the paper.

\begin{acknowledgement}
We thank Marcus T. Lexander for the discussions and assistance. 
This work was supported by the European Research Council (ERC) under the European Union's Horizon 2020 Research and Innovation Program (grant agreement No. 101020016). 
\end{acknowledgement}

\begin{suppinfo}
Details on the implementation, algorithm, 2D and circular scans and the expressions of the reduced and full space Hamiltonian matrices for both one and multiple projected states can be found in the Supporting Information. Furthermore, the ground and excited state energies of two non-interacting water molecules, thymine and thymine with a single helium atom when varying the number of projected states, and geometries, $\mathbf{g}$ and $\mathbf{h}$ vectors for all the systems considered.
\end{suppinfo}

\bibliography{bibliography}



\section*{Supplementary material}

# Supporting Information for "Generalized coupled cluster theory for ground and excited state intersections"

Federico Rossi, Eirik F. Kjønstad, Sara Angelico, and Henrik Koch*

*Department of Chemistry, Norwegian University of Science and Technology, NTNU, 7491 Trondheim, Norway*

E-mail: henrik.koch@ntnu.no

# Contents

# Implementation

A first version of the method was implemented in Julia, starting with a Hartree-Fock calculation using PySCF[1] to provide the integrals in the MO basis. The equations to construct omega and the product Jacobian times a vector was auto-generated using the Julia package `SpinAdaptedSecondQuantization.jl`[2] to provide the equations, which are then automatically converted into code based on np.einsums. A solver for the omega equations is implemented with DIIS acceleration to provide CCSD solutions, using `Arpack.jl` to iteratively solve the eigenvalue problem for the energies and vectors of the excited states. This code is used as a starting point to develop the GCCSD method, introducing the necessary projections, matrix construction and eigenvalue problem solvers. The results are tested on HeH$_2$ with STO-3G, to retrieve the same eigenvalues of an FCI reference calculation from eT.[3] For faster performance, a new version of the method is implemented in a local branch of eT, which is used for all the calculations reported in this work.

# Hamiltonian matrix in the new basis

## Single projected state

Reduced space Hamiltonian matrix

$$\mathbf{H}^{\mathrm{RS}} = \begin{pmatrix} \langle \mathrm{HF}|\bar{H}|\mathrm{HF}\rangle & \langle \mathrm{HF}|\bar{H}|R_1\rangle \\ \langle L_1|\bar{H}|\mathrm{HF}\rangle & \langle L_1|\bar{H}|R_1\rangle \end{pmatrix} = \begin{pmatrix} E_0 & \boldsymbol{\eta}^{\mathrm{T}}\mathbf{r}_1 \\ \mathbf{l}_1^{\mathrm{T}}\boldsymbol{\Omega} & E_0 + W_{11} \end{pmatrix} \quad (1)$$

where

$$W_{11} = \langle \mathrm{L}_1|\bar{H}|\mathrm{R}_1\rangle - E_0 = \omega_1 + \sum_{\nu}\langle \mathrm{L}_1|\mathrm{R}_1|\nu\rangle\Omega_\nu \quad (2)$$

Full space Hamiltonian matrix

$$\mathbf{H}^{\text{FS}} = \begin{pmatrix} \langle \text{HF}|\bar{H}|\text{HF}\rangle & \langle \text{HF}|\bar{H}|R_1\rangle & \langle \text{HF}|\bar{H}|\tilde{\nu}\rangle \\ \langle L_1|\bar{H}|\text{HF}\rangle & \langle L_1|\bar{H}|R_1\rangle & \langle L_1|\bar{H}|\tilde{\nu}\rangle \\ \langle \tilde{\mu}|\bar{H}|\text{HF}\rangle & \langle \tilde{\mu}|\bar{H}|R_1\rangle & \langle \tilde{\mu}|\bar{H}|\tilde{\nu}\rangle \end{pmatrix} = \begin{pmatrix} E_0 & \boldsymbol{\eta}^T \mathbf{r}_1 & X_\nu \\ \mathbf{l}_1^T \boldsymbol{\Omega} & E_0 + W_{11} & Y_\nu \\ 0 & V_\mu & Z_{\mu\nu} + \delta_{\mu\nu} E_0 \end{pmatrix} \quad (3)$$

where

$$X_\nu = \langle \text{HF}|\bar{H}|\tilde{\nu}\rangle = \eta_\nu - (\boldsymbol{\eta}^T \mathbf{r}_1) L_{1\nu} \quad (4)$$

$$Y_\nu = \langle L_1|\bar{H}|\tilde{\nu}\rangle = \langle L_1|\bar{H}|\nu\rangle - (E_0 + W_{11}) L_{1\nu} \quad (5)$$

$$V_\mu = \langle \tilde{\mu}|\bar{H}|R_1\rangle = \langle \mu|\bar{H}|R_1\rangle - (E_0 + W_{11}) R_{1\mu} \quad (6)$$

$$Z_{\mu\nu} = \langle \tilde{\mu}|\bar{H}|\tilde{\nu}\rangle - \delta_{\mu\nu} E_0 = \langle \mu|\bar{H}|\nu\rangle - \delta_{\mu\nu} E_0 - R_{1\mu} Y_\nu - (E_0 + W_{11}) R_{1\mu} L_{1\nu} - V_\mu L_{1\nu} \quad (7)$$

## Multiple projected states

Reduced space Hamiltonian matrix

$$\mathbf{H}^{\text{RS}} = \begin{pmatrix} \langle \text{HF}|\bar{H}|\text{HF}\rangle & \langle \text{HF}|\bar{H}|R_1\rangle & \ldots & \langle \text{HF}|\bar{H}|R_N\rangle \\ \langle L_1|\bar{H}|\text{HF}\rangle & \langle L_1|\bar{H}|R_1\rangle & \ldots & \langle L_1|\bar{H}|R_N\rangle \\ \vdots & \vdots & \ddots & \vdots \\ \langle L_N|\bar{H}|\text{HF}\rangle & \langle L_N|\bar{H}|R_1\rangle & \ldots & \langle L_N|\bar{H}|R_N\rangle \end{pmatrix} = \begin{pmatrix} E_0 & \boldsymbol{\eta}^T \mathbf{r}_1 & \ldots & \boldsymbol{\eta}^T \mathbf{r}_N \\ \mathbf{l}_1^T \boldsymbol{\Omega} & E_0 + W_{11} & \ldots & W_{1N} \\ \vdots & \vdots & \ddots & \vdots \\ \mathbf{l}_N^T \boldsymbol{\Omega} & W_{N1} & \ldots & E_0 + W_{NN} \end{pmatrix} \quad (8)$$

where

$$W_{IJ} = \langle L_I|\bar{H}|R_J\rangle - \delta_{IJ} E_0 = \delta_{IJ} \omega_I + \sum_\nu \langle L_I|R_J|\nu\rangle \Omega_\nu \quad (9)$$

Full space Hamiltonian matrix

$$\mathbf{H}^{\text{FS}} = \begin{pmatrix} \langle \text{HF}|\bar{H}|\text{HF}\rangle & \langle \text{HF}|\bar{H}|\text{R}_1\rangle & \ldots & \langle \text{HF}|\bar{H}|\text{R}_N\rangle & \langle \text{HF}|\bar{H}|\tilde{\nu}\rangle \\ \langle \text{L}_1|\bar{H}|\text{HF}\rangle & \langle \text{L}_1|\bar{H}|\text{R}_1\rangle & \ldots & \langle \text{L}_1|\bar{H}|\text{R}_N\rangle & \langle \text{L}_1|\bar{H}|\tilde{\nu}\rangle \\ \vdots & \vdots & \ddots & \vdots & \vdots \\ \langle \text{L}_N|\bar{H}|\text{HF}\rangle & \langle \text{L}_N|\bar{H}|\text{R}_1\rangle & \ldots & \langle \text{L}_N|\bar{H}|\text{R}_N\rangle & \langle \text{L}_N|\bar{H}|\tilde{\nu}\rangle \\ \langle \tilde{\mu}|\bar{H}|\text{HF}\rangle & \langle \tilde{\mu}|\bar{H}|\text{R}_1\rangle & \ldots & \langle \tilde{\mu}|\bar{H}|\text{R}_N\rangle & \langle \tilde{\mu}|\bar{H}|\tilde{\nu}\rangle \end{pmatrix} = \qquad (10)$$

$$= \begin{pmatrix} E_0 & \boldsymbol{\eta}^T \mathbf{r}_1 & \ldots & \boldsymbol{\eta}^T \mathbf{r}_N & X_\nu \\ \mathbf{l}_1^T \boldsymbol{\Omega} & E_0 + W_{11} & \ldots & W_{1N} & Y_{1,\nu} \\ \vdots & \vdots & \ddots & \vdots & \vdots \\ \mathbf{l}_N^T \boldsymbol{\Omega} & W_{N1} & \ldots & E_0 + W_{NN} & Y_{N,\nu} \\ 0 & V_{1,\mu} & \ldots & V_{N,\mu} & Z_{\mu\nu} + \delta_{\mu\nu} E_0 \end{pmatrix} \qquad (11)$$

where

$$X_\nu = \langle \text{HF}|\bar{H}|\tilde{\nu}\rangle = \eta_\nu - \sum_I (\boldsymbol{\eta}^T \mathbf{r}_I) \text{L}_{I\nu} \qquad (12)$$

$$Y_{I,\nu} = \langle \text{L}_I|\bar{H}|\tilde{\nu}\rangle = \langle \text{L}_I|\bar{H}|\nu\rangle - \sum_J (E_0 \delta_{IJ} + W_{IJ}) \text{L}_{J\nu} \qquad (13)$$

$$V_{I,\mu} = \langle \tilde{\mu}|\bar{H}|\text{R}_I\rangle = \langle \mu|\bar{H}|\text{R}_I\rangle - \sum_J \text{R}_{J\mu}(E_0 \delta_{IJ} + W_{JI}) \qquad (14)$$

$$Z_{\mu\nu} = \langle \tilde{\mu}|\bar{H}|\tilde{\nu}\rangle = \langle \mu|\bar{H}|\nu\rangle - \delta_{\mu\nu} E_0 - \sum_I \text{R}_{I\mu} Y_{I,\nu} - \sum_I V_{I,\mu} \text{L}_{I\nu} - \sum_{IJ} \text{R}_{I\mu}(E_0 \delta_{IJ} + W_{IJ}) \text{L}_{J\nu} \qquad (15)$$

## Algorithm and computational details

The full matrix eigenvalue equation in line 13 is solved using Davidson's algorithm. The initial vectors are set as:

$$\mathbf{x}_j = \begin{pmatrix} \mathbf{x}_j^{\text{RS}} \\ \mathbf{0} \end{pmatrix} \text{ for } j = 1, \ldots, n_{\text{proj}} + 1 \text{ and } \mathbf{x}_j = \begin{pmatrix} \mathbf{0} \\ \mathbf{R}_{j-1} \end{pmatrix} \text{ for } j = n_{\text{proj}} + 2, \ldots, n_{\text{excited}} + 1 \qquad (16)$$

**Algorithm 1** GCCSD algorithm
1: $k = 0$
2: $t^{[0]} =$ guess
3: **while** (k $<$ max$_{\text{iteration}}$ and $\|\tilde{\Omega}\|_{L_2} >$ threshold) **do**
4:     Solve the eigenvalue equation $\mathbf{A}\mathbf{r}_i^{[k]} = \omega_i^{[k]}\mathbf{r}_i^{[k]}, \quad i = 1, \ldots, n_{\text{excited}}$
5:     Solve the eigenvalue equation $\mathbf{A}^{\mathrm{T}}\mathbf{l}_i^{[k]} = \omega_i^{[k]}\mathbf{l}_i^{[k]}, \quad i = 1, .., n_{\text{excited}}$
6:     Biorthonormalize $\mathbf{r}_i^{[k]}$ and $\mathbf{l}_i^{[k]}$
7:     $t^{[k]} \leftarrow t^{[k]} - \hat{P}^{[k]}t^{[k]}$                                                  $\triangleright$ $\hat{P}^{[k]} = \sum_{i=1}^{n_{\text{proj}}} \mathbf{r}_i^{[k]}\mathbf{l}_i^{[k]T}$
8:     Construct $\Omega_\mu^{[k]}[t^{[k]}]$
9:     Remove the projection $\tilde{\Omega}_\mu^{[k]} = \Omega_\mu^{[k]} - \hat{P}^{[k]}\Omega_\mu^{[k]}$
10:    $t_\mu^{[k+1]} \leftarrow t_\mu^{[k]} - \tilde{\Omega}_\mu^{[k]}/\epsilon_\mu$                     $\triangleright$ $\epsilon_\mu$ is given by orbital energy differences
11:    $k \leftarrow k + 1$
12: **end while**
13: Construct and diagonalize the reduced matrix $\mathbf{H}^{\text{RS}}$
14: Solve the eigenvalue equation $\mathbf{H}^{\text{FS}}\mathbf{x}_j = \mathcal{E}_j\mathbf{x}_j, \quad j = 1, \ldots, n_{\text{excited}} + 1$

Note that at the beginning is not necessary to solve the eigenvalue problems in lines 4-5 with very tight threshold. To significantly reduce the number of iterations needed and the overall computational costs, the threshold used for the residual at the k-th macroiteration is set to be $\min(10^{-2}, 5\|\tilde{\Omega}^{[k]}\|_{L_2})$ so that it improves following the convergence of the cluster amplitudes.

Another issue to consider regards the overall sign of the eigenvectors, which is random after the diagonalization in the subspace of guess vectors. This does not change the projector, as left and right vectors are later biorthogonalized, but it makes following the phase of the eigenvectors difficult to follow. To avoid this when restarting a calculation from a previous geometry, the sign of each eigenvector is assigned in such a way that the dot product between the current eigenvector and the respective eigenvector at the previous geometry is positive. As long as the geometries are sufficiently similar, this ensures that the overall sign of the vector is continuous for neighboring geometries but still allows any single component of the vector to change sign, going through zero.

# Results on convergence and size extensivity

## Non-interacting identical water molecules

Table S1: Ground state and excited state CCSD and GCCSD energies for a single water molecule and two identical non-interacting water molecules. All calculations are performed with aug-cc-pVDZ, with the second water molecule translated by 500 Bohr on both $x$ and $z$. For a single water molecule, $\Delta$ $E_1$ refers to the first total symmetric state which is the third excited state and 3 states are included in the projector. For the case of two water molecules, $\Delta$ $E_1$ and $\Delta$ $E_2$ are the fifth and sixth excited states respectively and 6 states are included in the projector.

| System | $E_0/n_{H_2O}$ | $\Delta$ $E_1$ | $\Delta$ $E_2$ |
|---|---|---|---|
| CCSD 1 $H_2O$ | -76.269497284 | 0.347280380 | |
| GCCSD 1 $H_2O$ | -76.269497286 | 0.347201630 | |
| CCSD 2 $H_2O$ | -76.269497284 | 0.347280380 | 0.347280381 |
| GCCSD 2 $H_2O$ | -76.269497286 | 0.347201701 | 0.347239236 |

Table S2: Water geometries in Bohr.

| Atom | $x$ | $y$ | $z$ |
|---|---|---|---|
| O | 0.00000000 | 0.000000 | -0.009000 |
| H | 0.00000000 | 1.515263 | -1.058898 |
| H | 0.00000000 | -1.515263 | -1.058898 |
| O | 500.000000 | 0.000000 | -500.009000 |
| H | 500.000000 | 1.515263 | -501.058898 |
| H | 500.000000 | -1.515263 | -501.058898 |

# Effect of the number of states in the projector

Table S3: Thymine energies in Hartree when changing the number of projected states. All calculations are performed with cc-pVDZ, converged up to $1 \cdot 10^{-10}$. FS refers to eigenvalues of the full space matrix and RS to the one of the reduced space, with a number to indicate the number of states included in the projector.

| Method | $E_0$ | $E_1$ | $E_2$ | $E_3$ |
|---|---|---|---|---|
| CCSD | -452.9130722504 | -452.7700019437 | -452.7254480381 | -452.7002632744 |
| FS 1 | -452.9130722505 | -452.7700019491 | -452.7254479439 | -452.7002633747 |
| RS 1 | -452.9130722509 | -452.7700019484 | | |
| FS 2 | -452.9130738420 | -452.7708068743 | -452.7265615865 | -452.7008285885 |
| RS 2 | -452.9130960357 | -452.7708061397 | -452.7265005712 | |
| FS 3 | -452.9130787502 | -452.7711787685 | -452.7268942411 | -452.7006129360 |
| RS 3 | -452.9131081253 | -452.7711813076 | -452.7268576030 | -452.7005857256 |
| FS 4 | -452.9130783759 | -452.7711987402 | -452.7269036334 | -452.7006463304 |
| RS 4 | -452.9131083092 | -452.7712007102 | -452.7268674102 | -452.7006197380 |
| FS 5 | -452.9130795921 | -452.7711262607 | -452.7269319103 | -452.7005540834 |
| RS 5 | -452.9131103374 | -452.7711291875 | -452.7268949571 | -452.7005300811 |

Table S4: Thymine GCCSD excitation energy differences in eV when changing the number of projected states. The ground state energies in the first column are expressed in Hartree and expressed with respect to the CCSD value as $\Delta \text{GS} = E_{\text{method}}^{\text{GS}} - E_{\text{CCSD}}^{\text{GS}}$. All calculations are performed with cc-pVDZ, converged up to $1 \cdot 10^{-10}$.

| System | $\Delta$ GS | $\Delta E_1$ | $\Delta E_2$ | $\Delta E_3$ |
|---|---|---|---|---|
| CCSD | -452.9130722504 | 3.893141343 | 5.105514866 | 5.790827195 |
| GCCSD proj 1 | -1.1e-10 | 3.893141198 | 5.105517434 | 5.790824466 |
| GCCSD proj 2 | -1.6e-6 | 3.871281376 | 5.075256981 | 5.775487524 |
| GCCSD proj 3 | -6.5e-6 | 3.861295178 | 5.066338547 | 5.781489285 |
| GCCSD proj 4 | -6.1e-6 | 3.860741537 | 5.066072787 | 5.780570396 |
| GCCSD proj 5 | -7.3e-6 | 3.862746898 | 5.065336425 | 5.783113657 |

Table S5: Thymine and a single He atom shifted by 500 Å on all directions, when projecting different numbers of states. All calculations are performed with cc-pVDZ, converged up to $1 \cdot 10^{-10}$. FS refers to the eigenvalues of the full space matrix and RS to the one of the reduced space, with a number to indicate the number of states included in the projector.

| System | $E_0$-$E_{CCSD}^{He}$ | $E_1$-$E_{CCSD}^{He}$ | $E_2$-$E_{CCSD}^{He}$ | $E_3$-$E_{CCSD}^{He}$ |
|---|---|---|---|---|
| CCSD | -452.9130722504 | -452.7700019437 | -452.7254480381 | -452.7002632744 |
| FS 1 | -452.9130722505 | -452.7700019491 | -452.7254479439 | -452.7002633748 |
| RS 1 | -452.9130722509 | -452.7700019484 | | |
| FS 3 | -452.9130787502 | -452.7711787685 | -452.7268942411 | -452.7006129360 |
| RS 3 | -452.9131081253 | -452.7711813076 | -452.7268576030 | -452.7005857256 |
| FS 5 | -452.9130795921 | -452.7711262607 | -452.7269319103 | -452.7005540834 |
| RS 5 | -452.9131103374 | -452.7711291875 | -452.7268949571 | -452.7005300810 |

Table S6: Thymine geometry at the minimum in Angstrom.

| Atom | $x$ | $y$ | $z$ |
|---|---|---|---|
| C | 1.626856184467 | -0.090172437156 | 0.013282935761 |
| C | -0.197850164697 | 1.572388519191 | -0.033300408283 |
| C | -0.740709833470 | -0.739313238235 | -0.079521724593 |
| C | -1.176243220430 | 0.598835299691 | -0.065959144242 |
| C | -2.652953867469 | 0.918586250733 | -0.094848197888 |
| N | 0.630784739942 | -1.058855452370 | -0.174427947644 |
| N | 1.164641231138 | 1.205510451408 | -0.019672119665 |
| O | 2.794813959974 | -0.394624876859 | 0.158035521208 |
| O | -1.546761142159 | -1.810516595404 | -0.194583290615 |
| H | 1.881303447604 | 1.906943526328 | 0.115985843483 |
| H | -0.412611083215 | 2.640305248447 | -0.000321107921 |
| H | 0.924394117785 | -1.993233441610 | 0.093943333488 |
| H | -2.811463058403 | 2.006697839314 | -0.031363123742 |
| H | -3.180294992549 | 0.445499759315 | 0.751943675010 |
| H | -3.121321727307 | 0.554762000579 | -1.026259737912 |

# Ethylene

## Ethylene 2D scan

All 2D scans are run with the initial geometry, **g** and **h** vectors reported below. These vectors are originally determined in Hartree/Bohr and later used as displacement vectors in Bohr, defining a new geometry $\mathbf{r}_0 + \alpha\mathbf{g} + \beta\mathbf{h}$ from the initial geometry $\mathbf{r}_0$. For GCCSD the scheme used to cover the space was the one in Fig. S1a. For The CCSD results the scheme in Fig. S1b was chosen instead, to restrict the area where the flipped solution with negative excitation energy was obtained.

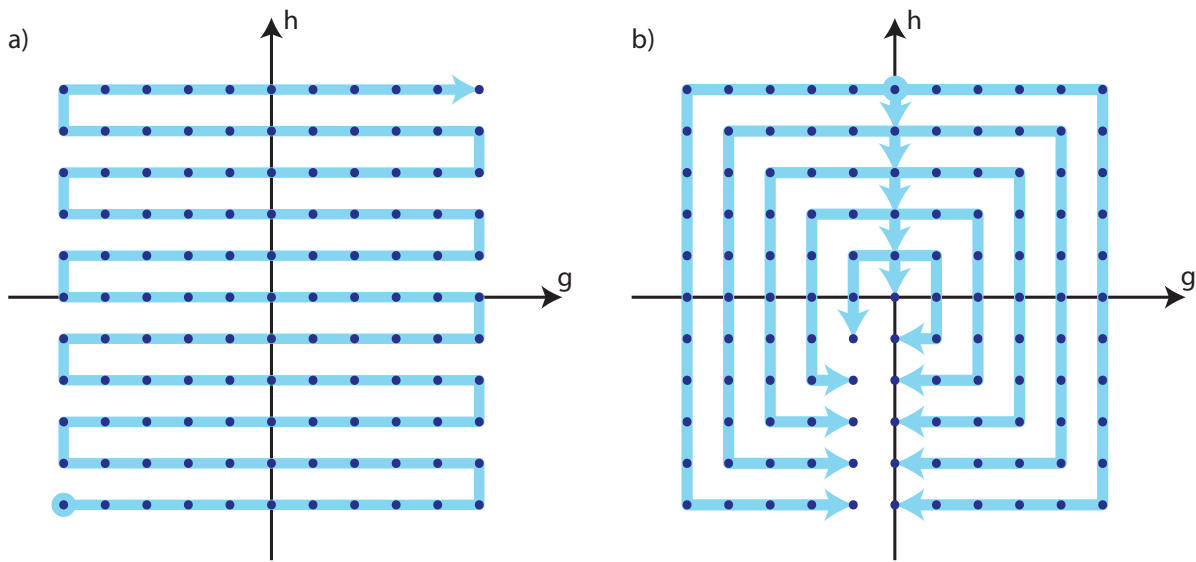

Figure S1: Graphical representation of the algorithms used to run the 2D scan.

Table S7: Ethylene $\varepsilon$-MECI geometry in Bohr.

| Atom | $x$ | $y$ | $z$ |
|------|-----|-----|-----|
| H | 1.999349425914 | -0.910907722158 | 1.414804072229 |
| C | 1.165469892760 | 0.249256850813 | -0.120813103612 |
| H | 2.556017310769 | 0.837765453705 | -1.559399709943 |
| C | -1.461989224355 | 0.680278349416 | -0.470086114356 |
| H | -2.617593462368 | 0.855685356721 | 1.279385377432 |
| H | -1.641442915322 | -1.523105689634 | -0.544646412150 |

Table S8: Ethylene g vector in Hartree/Bohr.

| Atom | $x$ | $y$ | $z$ |
|------|-----|-----|-----|
| H | 0.007313466289 | 0.057147259497 | 0.037233885635 |
| C | 0.018065045039 | 0.010653044239 | 0.029770694182 |
| H | -0.011045778402 | -0.044246241553 | -0.025640855466 |
| C | -0.016011637809 | -0.042775821534 | 0.029986440784 |
| H | 0.030212324793 | -0.002500610868 | 0.030496622594 |
| H | -0.028533419910 | 0.021722370219 | -0.10184678772] |

Table S9: Ethylene h vector in Hartree/Bohr.

| Atom | $x$ | $y$ | $z$ |
|------|-----|-----|-----|
| H | 0.009983729150 | 0.004279864574 | 0.006085526830 |
| C | -0.059128405254 | -0.031406366244 | 0.011698042309 |
| H | -0.000072261588 | -0.004346978757 | -0.005302673058 |
| C | 0.013554961398 | 0.069990614294 | -0.037047741578 |
| H | 0.008830292504 | -0.027241876753 | 0.014224759752 |
| H | 0.027122442072 | -0.011417198126 | 0.010473940680 |

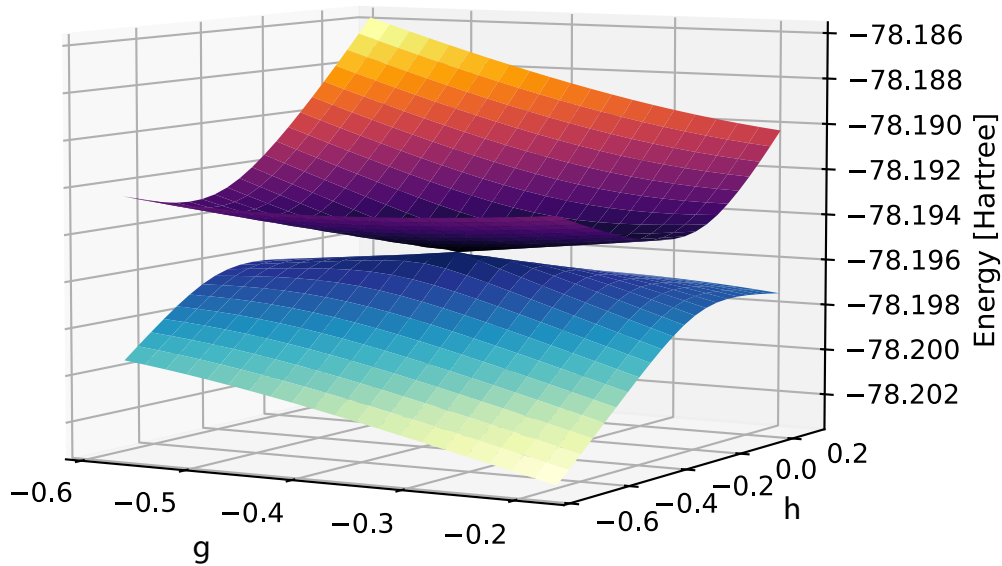

Figure S2: The GCCSD potential energy surfaces of $S_0$ and $S_1$ in ethylene. The basis is aug-cc-pVDZ and the energies are the total enegies expressed in Hartree.

## Ethylene circle

The **g** and **h** vectors are the same ones as in the previous section. The center of the circle is defined as $(g, h) = (-0.385, -0.13)$ from the $\varepsilon$-MECI, which is very close to the intersection for GCCSD, and the radius is radius $r = \sqrt{g^2 + h^2} = 0.8$. For the GCCSD calculations in

11

Fig. 4, the first point of the circle is (g, h) = (0.8, 0.0), set as 0°, and two full rotations are completed restarting each calculation from the previous geometry. For the CCSD results in Fig. 5a, the first point of the circle is (g, h) = (0.0, 0.8), corresponding to 0°, and the scan was run restarting for half a circle in one direction and then restarted from the initial point for half a circle in the opposite direction. The second CCSD circle in Fig. 5c was run following the same procedure but with the initial point in (g, h) = (0.0, −0.8).

## Ethylene extensivity and convergence

The structures used in Tables 1-2 as ethylene A, B, and C are obtained from the CCSD circle defined in the previous section, corresponding to the points at 0°, 3° and 6°. The corresponding geometries are reported below, with B and C translated by 1000 Å on $y$ and $z$ respectively.

Conversion factor: 1Å = 1.8897259886 Bohr

**Table S10: Ethylene A geometry in Angstrom.**

| Atom | $x$ | $y$ | $z$ |
|---|---|---|---|
| H | 1.0600599515526221 | -0.4921570182186516 | 0.7432539625898773 |
| C | 0.5920957988879612 | 0.11859558726990083 | -0.06584929946247757 |
| H | 1.3548109808694992 | 0.4507996323673442 | -0.8218550101479879 |
| C | -0.7655834382281663 | 0.39351776753311646 | -0.2680033422675551 |
| H | -1.3881953930153716 | 0.44366010709388964 | 0.6758518295627777 |
| H | -0.8531848120811245 | -0.8144664010539469 | -0.2637513912751824 |

**Table S11: Ethylene B geometry in Angstrom.**

| Atom | $x$ | $y$ | $z$ |
|---|---|---|---|
| H | 1.0598921220967552 | 999.5065743441498 | 0.7424254780268561 |
| C | 0.5917298546240902 | 1000.11837777963987291 | -0.0665156858139174 |
| H | 1.355055753036723 | 1000.4517824740106923 | -0.821283834929158 |
| C | -0.7652365486675872 | 1000.39442490172403727 | -0.2686462276042272 |
| H | -1.3888699003363099 | 1000.4437313156848888 | 0.6751678936041264 |
| H | -0.8525683614590025 | 999.1850589421331 | -0.26150095078339364 |

Table S12: Ethylene C geometry in Angstrom.

| Atom | $x$ | $y$ | $z$ |
|:---:|---:|---:|---:|
| H | 1.0597131680251948 | -0.4946957823769698 | -999.2584077970613 |
| C | 0.5914335230593863 | 0.11819701139648099 | -1000.06719381947529221 |
| H | 1.3552999381492163 | 0.45276766577214966 | -1000.8207080723028558 |
| C | -0.764906338412678 | 0.395248335882720 | -1000.2692443624734343 |
| H | -1.3895528051134067 | 0.44383393922191544 | -999.32553067344 |
| H | -0.8519850720229274 | -0.815401165727728 | -1000.2592688320343005 |

**Table S13:** A detailed breakdown of the individual contributions to the total time for the case of an ethylene molecule, as described in Table 2. The timings are wall times in seconds on an Intel Xeon Gold 6342 using 24 cores. For GCCSD, the projection of the omega, the projection of the amplitudes and the construction and diagonalization of the reduced matrix are all included in the Ground state section.

| CCSD | Section | GCCSD |
|---|---|---|
| 1.76s | Reference wavefunction | 1.56s |
| 10.94s | Cholesky decomposition | 6.51s |
| 9.61s | Ground state | 9.72s |
| 16.78s | Right eigenvectors | 32.88s |
| 17.63s | Left eigenvectors | 33.24s |
| - | Full matrix eigenvectors | 12.22s |
| 56.96s | Total time | 96.33s |

For CCSD, the number of iterations to determine the right eigenvectors of the Jacobian was 33, for a total of 87 Jacobian transformations. The number of iterations to determine the left eigenvectors of the Jacobian was 33, for a total of 84 Jacobian transpose transformations.

For GCCSD, the number of iterations to determine the right eigenvectors of the Jacobian was 101, for a total of 244 Jacobian transformations. The number of iterations to determine the left eigenvectors of the Jacobian was 104, for a total of 231 Jacobian transpose transformations. The number of iterations to determine the right eigenvectors of the full matrix was 27, for a total of 69 full matrix transformations.

We point out that for GCCSD all 3 right and left eigenvectors are determined and updated when solving for the ground state, even though only one state is projected. This was done to show a fair comparison between the methods, as all 3 states are required for CCSD. In GCCSD, only the eigenvectors that need to be projected are required during the ground state procedure, resulting in a decrease in computational cost. All the remaining eigenvalues and eigenvectors are later obtained through the final solution of the full matrix eigenvalue problem. We also note that this final step of the algorithm presents a computational cost and scaling with system size that is nearly identical to solving the right eigenvalue problem for the Jacobian in standard CCSD, since the complexity of the contributions is identical

and the size of the problem is increased only by the number of projected states.

# Thymine 2D scan

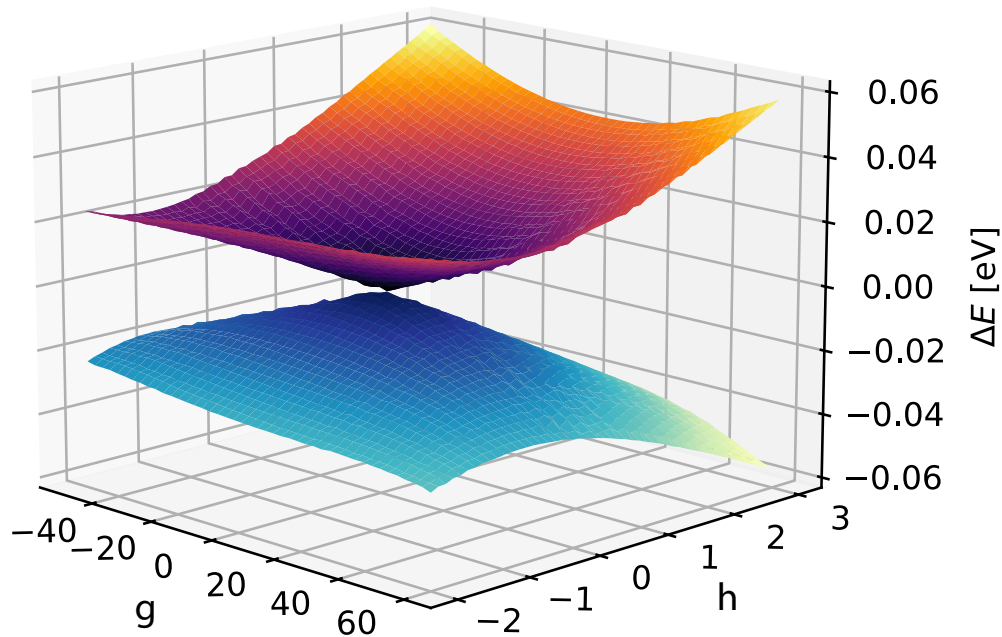

Figure S3: SCCSD potential energy surfaces of $S_1$ and $S_2$ in thymine with cc-pVDZ basis. All energies are plotted in eV, relative to the average $\frac{1}{2}(E_1 + E_2)$ for each point. The plot shows the same region of Fig. 3c in Ref. 4, which is larger than what shown in Fig. 6. The axes have also been rescaled to be consistent with Fig. 6. Note that the thresholds for the calculations used in the reference are different, which results in visible noise.

Table S14: Thymine initial geometry in Bohr.

| Atom | $x$ | $y$ | $z$ |
|---|---|---|---|
| C | 3.024632508678 | -0.197939565118 | -0.043506106319 |
| C | -0.379163249776 | 2.993201825994 | -0.376843901976 |
| C | -1.584772946934 | -1.537050600849 | -0.174652122376 |
| C | -2.22289716323 | 1.092687974969 | -0.019993160668 |
| C | -4.942441756804 | 1.880988047973 | 0.295884973444 |
| N | 1.383965794231 | -2.10204230652 | -0.134678067251 |
| N | 2.112941375643 | 2.387375342191 | 0.152606749372 |
| O | 5.327286048654 | -0.549667905189 | 0.112112165406 |
| O | -2.956702928079 | -3.541736963203 | 0.113404299659 |
| H | 3.56372918388 | 3.213977204429 | 0.959418384849 |
| H | -1.197624746346 | 4.870276855875 | -1.007782429014 |
| H | 2.125362704196 | -3.672844271999 | -0.139754412083 |
| H | -5.173960594864 | 3.777158444758 | 1.238505943709 |
| H | -6.247239158792 | 0.021538888608 | 0.656427604651 |
| H | -5.484297294198 | 2.089178080965 | -1.738950890195 |

Table S15: Thymine g vector in Hartree/Bohr.

| Atom | $x$ | $y$ | $z$ |
|---|---|---|---|
| C | -0.0005798127253398536 | -0.0006105588384348458 | -0.0003687934628399069 |
| C | 0.0003818164857299036 | -0.0002872196909549275 | -0.00020855092277994734 |
| C | -0.000807956004604796 | 0.0004760994383148798 | -0.0002163023882049454 |
| C | -0.00026859391788993216 | 0.00032664710958991755 | -8.684068250997807e-05 |
| C | 0.0004941927063798752 | 0.0008342066308997893 | 0.0006916614546198255 |
| N | 0.0024098330181643915 | -0.002301077258679419 | 0.00022497031079994322 |
| N | 0.00026493162539493315 | 0.0012347579959946884 | 0.0005719552842648555 |
| O | 0.00026468183472493317 | -0.00011783293740497025 | 5.885903984998514e-05 |
| O | -0.00025493002853493565 | -0.00037689204538490486 | 2.6315375574993355e-05 |
| H | -0.0003681107651799071 | -0.0008153142290797942 | -0.00019227043060995145 |
| H | -0.00051518556615487 | 0.0024671446873993774 | -0.00016483332544495842 |
| H | -0.0007069723990748215 | 0.0021578015843344554 | 4.683741952998818e-05 |
| H | 0.0001180527907449702 | 0.00046875080507488163 | -5.661090652498571e-05 |
| H | -0.0005499230343048612 | -0.0009651818081297563 | -0.00026794109715993234 |
| H | 0.00011797597996497022 | -0.0002709012248249316 | -5.8455668569985244e-05 |

Table S16: Thymine h vector in Hartree/Bohr.

| Atom | $x$ | $y$ | $z$ |
|---|---|---|---|
| C | 0.00021457107071499082 | 0.001803226261795447 | 9.749523596998285e-05 |
| C | -0.0010134839562702467 | -0.0009720296212803169 | 0.0004611740828001079 |
| C | 0.0016994394176503926 | 0.0028170744749808733 | -0.0003725336787013453 |
| C | 0.00038991867326007985 | 0.0004750303736451777 | -6.875357674503058e-05 |
| C | -0.0002448460136600101 | 4.055406735511435e-05 | -0.0004715577597300515 |
| N | -0.00039165567384481694 | -0.0009536981321605591 | 5.521516118504366e-05 |
| N | 0.001120233510195357 | -0.0005743336152900145 | -0.000284505319725012 |
| O | -0.0002510358241650401 | -0.0005482676678901734 | -2.513796076000004e-05 |
| O | -0.0009858319883903168 | -0.001872929150235589 | 0.00039741206067511834 |
| H | -0.00022507579131511045 | -0.0001022527671401299 | -7.556058939004552e-05 |
| H | -0.00010585761811509402 | -6.75180585849892e-05 | 0.00013027901358001745 |
| H | -1.3309690020090808e-05 | -7.338529004736723e-06 | 8.97661201500836e-06 |
| H | 1.620964079001921e-05 | -8.850160360996796e-05 | 3.125465966500208e-05 |
| H | -8.820500490509318e-05 | 5.681394686489774e-05 | 1.355151105497097e-05 |
| H | -0.00012105746345002055 | -5.786971995034995e-06 | 0.00010269838165002255 |

## 2,4-Cyclohexadien-1-ylamine 2D scan

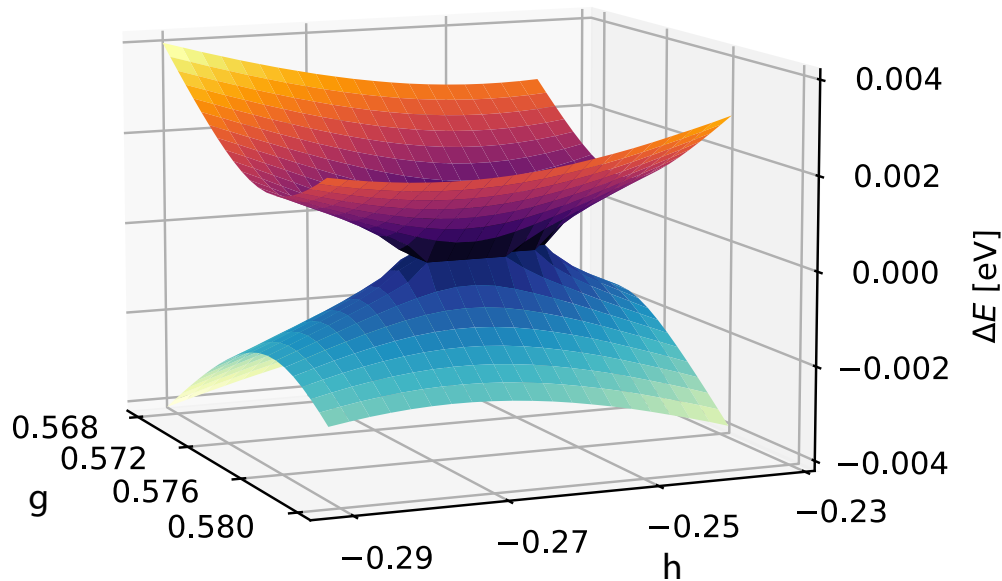

Figure S4: The GCCSD potential energy surfaces of $S_0$ and $S_1$ in 2,4-cyclohexadien-1-ylamine with cc-pVDZ in the region very close to the intersection. The energies for each point are plotted in eV relative to the average energy $\frac{1}{2}(E_0 + E_1)$.

Table S17: 2,4-Cyclohexadien-1-ylamine initial geometry in Bohr.

| Atom | $x$ | $y$ | $z$ |
| --- | --- | --- | --- |
| N | 2.485229105603 | 0.547832318017 | -1.016626362773 |
| C | -1.020180309090 | -2.514973398583 | -1.130137584081 |
| C | 0.000000000000 | 0.000000000000 | 0.000000000000 |
| C | -0.152827830887 | -4.842609039125 | 0.243271218683 |
| C | 0.000000000000 | 0.000000000000 | 2.948948714717 |
| C | 0.393673950618 | -4.812906418385 | 2.772429176890 |
| C | 0.638744772120 | -2.397590397164 | 4.027981665337 |
| H | 3.068814728580 | 2.219087393160 | -0.340105507921 |
| H | 3.743103862610 | -0.735412398662 | -0.407797247778 |
| H | -0.597826785001 | -2.656733416686 | -3.131827502962 |
| H | -1.241628965952 | 1.485927944692 | -0.652347722798 |
| H | -3.073407175968 | -2.431583674518 | -0.951269929831 |
| H | -0.051896980885 | -6.598968101875 | -0.785039323552 |
| H | -2.020803718174 | -1.042080601584 | 3.045064771778 |
| H | 0.977265443085 | -6.512166056261 | 3.733568986966 |
| H | 1.430957011893 | -2.465032088650 | 5.918242701070 |

Table S18: 2,4-Cyclohexadien-1-ylamine g vector in Hartree/Bohr.

| Atom | $x$ | $y$ | $z$ |
|---|---|---|---|
| N | 0.004971062452 | -0.000006312063 | -0.008307843382 |
| C | -0.016216875814 | -0.011028497968 | 0.013811359333 |
| C | 0.028843656521 | 0.052451171540 | 0.008092929065 |
| C | -0.002450100320 | -0.007281793260 | -0.010822870907 |
| C | -0.081339063409 | -0.074869115270 | -0.049735972788 |
| C | -0.013448003251 | 0.032461991297 | 0.014876665117 |
| C | 0.068149686093 | -0.043067069541 | 0.067549127712 |
| H | 0.000405335195 | -0.001041323709 | -0.002325561626 |
| H | -0.001324788614 | -0.000710206799 | -0.000031584236 |
| H | 0.003448251125 | 0.006602895615 | 0.002116067133 |
| H | -0.001001097065 | -0.000044020470 | -0.002328844295 |
| H | 0.003064563939 | 0.003521849557 | -0.003645421539 |
| H | 0.001814047671 | 0.001543968627 | -0.001141347406 |
| H | -0.009685201838 | 0.038125491323 | -0.021492693576 |
| H | -0.000178824202 | 0.003941613764 | 0.002797133980 |
| H | 0.014947351518 | -0.000600642643 | -0.009411142586 |

Table S19: 2,4-Cyclohexadien-1-ylamine h vector in Hartree/Bohr.

| Atom | $x$ | $y$ | $z$ |
|---|---|---|---|
| N | 0.002487591355 | 0.001059423611 | -0.001229774795 |
| C | 0.001417281298 | 0.007961593644 | 0.001946831613 |
| C | 0.002788554845 | -0.012082019030 | -0.001934955082 |
| C | 0.000037767487 | -0.003593455041 | 0.002688939505 |
| C | -0.010315266883 | -0.006727478406 | -0.008749189841 |
| C | 0.018041726684 | 0.000359400262 | -0.009740084068 |
| C | 0.007161006545 | -0.003086550689 | -0.006418446171 |
| H | 0.000523768377 | 0.000163090505 | 0.000341198029 |
| H | 0.000201834734 | 0.000043417177 | -0.000047450886 |
| H | 0.000287627165 | 0.001485980656 | 0.000635812663 |
| H | 0.000112190184 | 0.000739350784 | 0.000996160267 |
| H | -0.001339158720 | -0.002603931746 | -0.001285345959 |
| H | -0.003490679658 | -0.000662035975 | 0.000729035443 |
| H | -0.004741496715 | 0.016752015594 | 0.017562156627 |
| H | -0.001039863972 | -0.000167499259 | 0.000074916208 |
| H | -0.012149841791 | 0.000529111955 | 0.004488870984 |




\end{document}